\documentclass[a4paper,fleqn,usenatbib]{mnras}

\usepackage{newtxtext,newtxmath}

\usepackage[T1]{fontenc}
\usepackage{ae,aecompl}

\usepackage{graphicx}	
\usepackage{amsmath}	
\usepackage{amssymb}	
\usepackage{caption}
\usepackage{subfigure}

\title[Truncations above the galaxies' mid-plane]{Discovery of disc truncations above the galaxies' mid-plane in Milky Way-like galaxies}

\author[C. Mart{\'i}nez-Lombilla, I. Trujillo \& J. H. Knapen]{
Cristina Mart{\'i}nez-Lombilla,$^{1,2}$\thanks{E-mail: cml@iac.es}
Ignacio Trujillo$^{1,2}$
and Johan H. Knapen$^{1,2,3}$ 
\\
$^{1}$Instituto de Astrof{\'i}sica de Canarias (IAC), La Laguna, 38205, Spain\\
$^{2}$Departamento de Astrof{\'i}sica, Universidad de La Laguna (ULL), E-38200, La Laguna, Spain\\
$^{3}$Astrophysics Research Institute, Liverpool John Moores University, IC2, Liverpool Science Park, 146 Brownlow Hill, Liverpool, L3 5RF, UK\\
}

\date{Accepted XXX. Received YYY; in original form ZZZ}

\pubyear{2018}

\begin{document}
\label{firstpage}
\pagerange{\pageref{firstpage}--\pageref{lastpage}}
\maketitle

\begin{abstract}

Disc truncations are the closest feature to an edge that galaxies have, but the nature of this phenomenon is
not yet understood. In this paper we explore the truncations in two nearby ($D\sim$15~Mpc) Milky Way-like galaxies: NGC~4565 and NGC~5907. We cover a wide wavelength range from the NUV and optical to $\mathrm{3.6 \, \micron}$. We find that the radius of the truncation (26 $\pm$ 0.5~kpc) is independent of wavelength. 
Surprisingly, we identify (at all wavelengths) the truncation at altitudes as high as 3~kpc above the mid-plane, which implies that the
thin disc in those outer regions has a width of at least this value. We find the
characteristic U shaped radial colour profile associated with a star formation threshold at the location of the
truncation. Further supporting such an origin, the stellar mass density at the position of the truncation is
$\sim$1--2~$M_{\odot}$~pc$^{-2}$, in good agreement with the critical gas density for
transforming gas into stars. Beyond the truncation, the stellar mass in the mid-plane of the
disc drops to just 0.1--0.2\,\% of the total stellar mass of the galaxies. The detection of the truncation at high altitude in combination with the U shape of the radial colour profile allows us to establish, for the first time, an upper limit to the present-day growth rate of
galactic discs. We find that, if the discs of these galaxies are growing inside-out, their growth rate is less than 0.6--1~kpc~Gyr$^{-1}$.

\end{abstract}





\section{Introduction} \label{Intro}

Many discs of spiral galaxies have a well-defined feature in their periphery that is called a
truncation. Truncations are found in three out of four galaxy thin discs
\citep{KruitFreeman2011, Comeron2012} and typically occur at radial distances of
four or five times the exponential scale length of the inner disc. Beyond the truncation the light
distribution declines rapidly
\citep{vanderKruit1979,vanderKruit1981a,vanderKruit1981b}. The origin of this
effective frontier for disc galaxies is yet unclear. Some authors suggest it could be
linked to a threshold in star formation \citep{Kennicutt1989,Rovskar2008b,
Rovskar2008a}, for others, the truncation reflects the location of those stars with the largest angular momentum at the
moment of the collapse of the protogalaxy \citep[][]{VanDerKruit1987}. Other
alternatives in the literature to explain the origin of breaks in the surface brightness
distribution of the discs include bar angular momentum distribution \citep{Debattista2006}, or any other driver such as density waves, or the heating and stripping of stars by bombardment of dark matter subhaloes \citep[e.g.,][]{LaurikainenSalo2001,deJong2007}. In order to probe the nature of this edge, we explore whether its location depends on either the wavelength at
which it is measured or the height above the galaxy mid-plane \citep[see e.g.,][]{deJong2007}. The truncation of discs can be seen at any galaxy
orientation but, due to its location in the outer disc (and consequently at low surface
brightness), an edge-on orientation is the most favoured one \citep[see discussion in][]{Martin-Navarro2012}.

Truncations should not be confused with the surface brightness breaks in Type II surface brightness profiles found at smaller radii \citep[roughly two times closer to the galaxy centre, as reported in the unified picture of breaks and truncations in spiral galaxies proposed by][]{Martin-Navarro2012}. This means that the features labelled as truncations by \cite{PohlenTrujillo2006} most probably are disc breaks rather than the face-on counterparts of truncations observed in edge-on galaxies. In the review by \cite{Debattista2017}, breaks are defined as a break in the star formation, not as the bumps and wiggles produced by bars, rings, or spirals inside the disc, while truncations imply a sharp edge. In this work we adopt the assumption of \cite{Martin-Navarro2012}, namely that type II breaks correspond to a feature in galaxy discs (mainly connected to a break in star formation) at closer radial distances than the truncations and that, for many (but not all) galaxies, breaks and truncations can coexist as two differentiated features. The ability to detect disc truncations is highly conditioned by the galaxy's orientation. In face-on systems the truncations can be seen only in very deep surface photometry, as they are fainter than in the edge-on view \citep[e.g.,][]{Martin-Navarro2012,Peters2017}. This is because of the reduced line-of-sight integration through the disc. Nevertheless, truncations can also be hidden by the lopsided nature of spiral galaxies \citep[e.g.,][]{Zaritsky2013}, and/or by light from a faint stellar halo \citep{Pohlen2002,Martin-Navarro2012} and/or from the inner disc scattered by the PSF \citep{deJong2008,Sandin2014,Sandin2015}.

If truncations are connected with an inside-out propagation of star formation in the
disc, the expectation is that the location of this feature changes as cosmic time
progresses (such cosmic evolution has been observed for the location of type II
breaks in disc galaxies since $z\sim$1, \citealt{TrujilloPohlen2005,Azzollini2008}).
If this were the case, the younger populations would be expected to show a truncation at
larger radial distances than older populations. Similarly, due to the time delay for the
stellar populations originated in the plane of the disc to reach a given distance above the
galaxy mid-plane, the location of the truncation at large heights would be expected to occur at smaller radial distances than for stars at the mid-plane. This change in the position of the truncation would depend on its stellar population ages and so its radial position would change gradually. Consequently, one could use the location of the radial truncation of discs to measure ongoing
growth (if any) of disc galaxies. We explore these ideas in this paper.

If breaks and truncations had a common origin, predictions in \cite{Rovskar2008a} might be relevant. They propose the origin of breaks in an inside-out disc growth scenario, where the stellar disc is formed with a surface density profile consisting of an inner exponential breaking to a steeper outer exponential. The break forms early on and persists in time, moving outward as more gas is able to cool and add mass to the disc. This break is associated to a drop in the cooled gas surface density (i.e., a radial star formation cut-off), while the stellar populations in the outer exponential were scattered outward on nearly circular orbits from the inner disc by spiral arms (i.e., secular evolution processes). A consequence of such a formation and evolution scenario is a sharp change in the radial mean stellar age profile at the break radius, which is in excellent agreement with the observations reported in \cite{Bakos2008}. \cite{Rovskar2008a} suggest that the location of the break radius evolves with time (see their Fig.~1), while it is independent of the age and of height (see their Fig 3). The simulations thus show that migration forces a common break radius for all stellar populations, even ones born at a time when the break radius was much smaller. We explore this issue here.

Galaxies grow at different speed depending on their mass \citep[see
e.g.,][]{Trujillo2004,Trujillo2006}. To facilitate the
interpretation of the data, we have thus explored two edge-on disc galaxies with masses similar
to the Milky Way: NGC~4565 and NGC~5907 ($v_{\rm rot} \sim 230$~km s$^-1$). To probe the ongoing
growth of the galaxies it is necessary to have enough spatial and time resolution. The closer the
galaxies are, the better the spatial resolution to explore the locations
of the truncation as a function of wavelength and/or height. For this reason, we have
taken two galaxies that are at a distance less than 20~Mpc. At the typical 1~arcsec
resolution from ground-based data, this implies that the minimum difference in spatial
location we can resolve is $\sim$\,100~pc. To probe whether the truncation is moving (i.e., to measure its radial velocity), it is also necessary to have a time estimator. Throughout this work, we will use two different clocks. For the first approach, we use as a proxy
to measure time the change in the NUV-\textit{r} colour of recently formed stellar populations. We
will show that intervals of time of just a few hundred Myr can produce variations of such a
colour of $\sim$1~mag. The other time indicator we will use is the time that it takes for a star
originated in the disc plane to reach a certain height. We will show that at the radial
distance of the truncation, the time that a star needs to arrive a few~kpc above the disc is
(again) a few hundred Myr. These two time indicators, together with our spatial
resolution, allow us to measure growth rates with an accuracy better than 0.5~kpc Gyr$^-1$.

Finally, in addition to measuring the growth rate of the disc, finding disc
truncations at different heights above the galaxy mid-plane can help to measure the
``thickness" of the thin disc. Although the origin of the truncation is not clear, if indeed it
is related to a star formation threshold, then the truncation is expected to be a
phenomenon linked exclusively with the thin disc. In this sense, if the truncations are
clearly visible up to a given height, we could claim with some confidence that at such a
radial location the thin disc is still present. Consequently, the highest
location at which the disc truncation can be measured can be used as a measure of the
vertical extent of the thin disc.

The paper is structured as follows. In Sect.~\ref{section:SelectionData} we present our sample, selection criteria, and data. Then, we describe in detail the method and how the profiles are extracted (Sect.~\ref{section:Method}). The results are shown in
Sect.~\ref{section:results} and analysed in Sections~\ref{section:growth}, \ref{section:comologicalGrowth} and \ref{section:physicalOrig}. All magnitudes are provided in the AB system.

\section{Target selection and data}  \label{section:SelectionData}

\begin{figure*}
\includegraphics[width=\textwidth]{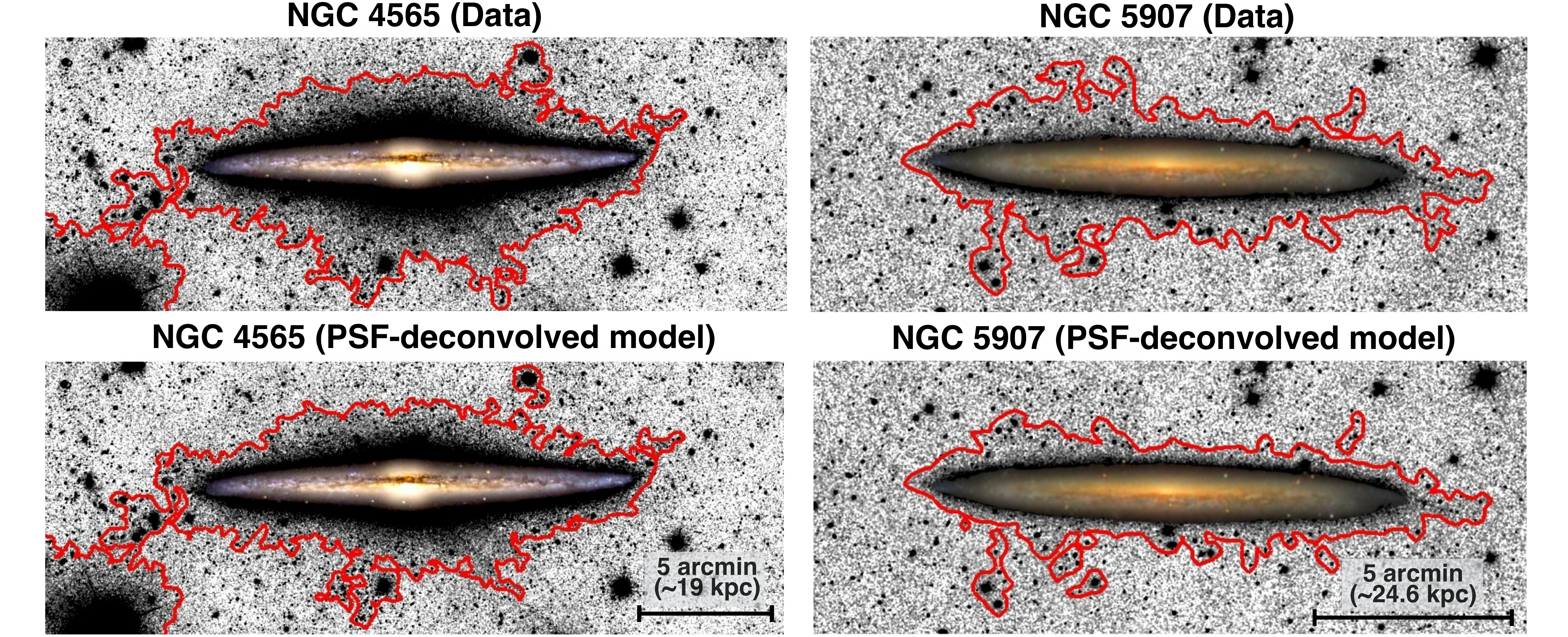}

\caption{Images of NGC~5907 (left) and NGC~4565 (right) in SDSS \textit{gri} combined light
(roughly equivalent to a deep \textit{r}-band image). For illustrative purposes, a colour image obtained with
the same telescope has been inserted atop the saturated disc region of the galaxy. Panels show in greyscale and for each galaxy the observed data, in the case of the first row, and the PSF-deconvolved models in the second row (see Sect.~\ref{subsection:PSF} for details). The colour image of each galaxy is the same for both PSF-convolved and PSF-deconvolved models. Overplotted on all the images are the surface brightness contours (red) corresponding to 27~mag~arcsec$^{-2}$. These contours are located closer to the galaxy's mid-plane when the scattered light is corrected.}

\label{fig:GalColourComposition}
\end{figure*}

We study the position of the truncation as a function of the height above the mid-plane in
the two well-known highly inclined nearby galaxies NGC~4565 and NGC~5907 (see
Fig.~\ref{fig:GalColourComposition}). We cover a wide wavelength range, from the near
ultraviolet (NUV) and optical, to the infrared (IR). Table~\ref{tabla:paramfisicos}
summarizes the general information and physical parameters of both galaxies. For each
object, the Hubble type was taken from the \textit{Spitzer} Survey of Stellar Structure in
Galaxies \citep[S$^{4}$G,][]{Sheth2010} morphological classification of \cite{Buta2015} and the distance
was obtained from the most recent values in the NASA$/$IPAC Extragalactic
Database\footnote{http://ned.ipac.caltech.edu/} (NED). The
other properties were taken from the HyperLeda\footnote{http://leda.univ-lyon1.fr/} 
database \citep{Makarov2014}.

\subsection{Selection criteria} \label{subsection:selectionCriteria}

The galaxies were selected based on their apparent size, mass, inclination, morphological
type and on the availability of images in the wide spectral range considered. In particular,
we need galaxies with large apparent sizes ($d_{25}\gtrsim 10$~arcmin) in order to reach
good physical spatial resolution. We also want very inclined galaxies (incl\,$\gtrsim
85$~deg) to ensure that our analysis is done in the vertical direction of the object. We
choose Milky Way-like galaxies in terms of rotational velocity \citep[$v_{\rm rot}\sim230$~km s$^-1$,][]{BlandHawthorn2016} but with different morphology (see Table
\ref{tabla:paramfisicos}). The main reason is that several N-body simulations have shown relations between the
sources of internal secular vertical heating and galaxy morphology, such as spiral arms
or bars \citep[e.g.,][]{Sellwood1984, MinchevQuillen2006, Grand2016}. In
this case, the main difference is the presence of a bar in NGC~4565 but not in NGC~5907 \citep{Buta2015}.
Finally, we checked Sloan Digital Sky Survey Data Release 12 (SDSS DR12) images
\citep{Alam2015} to avoid galaxies with bright foreground stars or very strong
dust lanes. 

\begin{table*}
\centering
\caption{Properties and physical parameters of NGC~4565 and NGC~5907. Columns represent: (1) galaxy ID; (2) morphological type from \citet{Buta2015}; (3) distance in Mpc from the average of the most recent values (since 2010) in the NED; (4) major axis position angle (North Eastwards) from the HyperLeda database \citep{Makarov2014}; (5) absolute \textit{B}-band magnitude \citep[HyperLeda,][]{Makarov2014}; (6) maximum rotational velocity corrected for inclination \citep[HyperLeda,][]{Makarov2014}; (7) galaxy diameter in arcmin at the isophotal level 25~mag~arcsec$^{-2}$ in the \textit{B}-band corrected for galactic extinction \citep[HyperLeda,][]{Makarov2014}; and finally, (8) inclination to the line-of-sight in degrees obtained by the authors of this work assuming that the most dense part of the dust lanes of the galaxies are located in the mid-plane.} 
\label{tabla:paramfisicos} 
\begin{tabular}{c||c|c|c|c|c|c|c}
\hline 
Galaxy & Morph. & $D$ & PA & $M_{abs}$ & $v_{\rm rot}$ & $d_{25}$ & Incl. \\  
 ID &type&[Mpc]&[deg.]& [\textit{B}-mag] & [km s$^-1$] & [arcmin] & [deg] \\
\hline 
NGC~4565 & SB$_{\rm {X}}$(r)\underline{a}b spw & 13.1 $\pm$ 0.5 & 135.2 & -21.44 $\pm$ 0.24 & 243.6 $\pm$ 4.7 & 16.6 $\pm$ 0.1 & 88.5 $\pm$ 0.5\\ 
 
NGC~5907 & SA:(nd)\underline{c}d spw & 16.9 $\pm$ 0.3 & 155.6  & -21.48 $\pm$ 0.14 & 	226.6 $\pm$ 4.3 & 	11.2 $\pm$ 0.1 & 87.5 $\pm$ 0.5 \\ 

\hline 
\end{tabular}
\end{table*}

\subsection{\textit{GALEX} NUV data} \label{subsection:NUV}

The ultraviolet data were taken from the \textit{Galaxy Evolution Explorer}
\citep[\textit{GALEX},][]{Martin2005, Morrissey2007} Guest Investigator
Program (GIP, PI V. Buat; NGC~4565: GI6-049003, NGC~5907: GI6-049002) in the NUV
band (1750-2800$\,{\rm \AA}$, 2267$\,{\rm \AA}$ effective wavelength). \textit{GALEX} images have a
circular field of view (FOV) of 1.2~degrees, and a spatial resolution (FWHM) of 5.3~arcsec in
the NUV channel sampled with a pixel size of 1.5~arcsec. The exposure times of our
images are 12050.15 seconds in fourteen visits for NGC~4565 and 15118.45 seconds in eleven visits for
NGC~5907, which gives a surface brightness limit of $\mu _{AB}\sim30.5$~mag~arcsec$^{-2}$ (1$\sigma$). To convert intensities to magnitudes, we used the \textit{GALEX} zero point magnitude from \cite{Morrissey2007}, $m_{0, NUV} = 20.96$.

\subsection{SDSS optical data} \label{subsection:SDSS}

We use optical data from the SDSS DR12 \citep{Alam2015} in the \textit{g}, \textit{r} and \textit{i}
bands. As we are studying extended objects, we used the SDSS mosaic
tool\footnote{http://dr12.sdss.org/mosaics/}, which stitches together the corrected frames in a selected region to form coherent images over larger patches of the sky. These mosaics were created using the SWARP \citep{Bertin2002} utility. The resulting
images were calibrated and sky-subtracted just as the corrected frames themselves are, but
we also estimate our sky background independently (see Sect.~\ref{subsection:SkySub} for
more details).

The SDSS survey images have a pixel size of 0.396~arcsec and a mean PSF FWHM value in the
\textit{r}-band of 1.3~arcsec. The exposure time for all the images is 53.9 seconds. After combining the \textit{g}, \textit{r} and \textit{i} bands by adding them all together, in order to reach a deeper surface brightness limit, we could to extract reliable surface brightness profiles down to $\mu \sim 27.0$~mag~arcsec$^{-2}$ with a fixed zero point value\footnote{http://www.sdss.org/dr12/algorithms/magnitudes/} of $m_{0,gri} = 20.49$. We consider a profile reliable if by using $\pm1\sigma_{\rm sky}$ in the profile extraction, it deviates less than 0.2~mag from the one with the mean sky value \citep[][]{PohlenTrujillo2006}. The combined \textit{gri} profile is roughly equivalent to the \textit{r}-band profile: we did not find any departure of the \textit{gri} combined data from the \textit{r}-band larger than 0.1~mag along the entire radial profile in the region where the photometric uncertainty is negligible.

\subsection{Infrared S$^{4}$G data} \label{subsection:IR}

The S$^{4}$G project \citep{Sheth2010} obtained 3.6 and 4.5~$\micron$ images of
2352 nearby galaxies ($D < 40$~Mpc) using the Infrared Array Camera
\citep[IRAC,][]{Fazio2004}. We are particularly interested in the 3.6~$\micron$
band (IRAC channel 1) because it is a reasonably good tracer of stellar mass density and the images are less affected by dust. This
is a key point when dealing with edge-on galaxies as there is a lot of dust along the line
of sight. The S$^{4}$G team have reduced all the data using their pipelines
\citep{Munoz-Mateos2015, Querejeta2015, Salo2015}. 

The S$^{4}$G IRAC 3.6~$\micron$ images have a pixel size of 0.75~arcsec and each galaxy
was observed for 240 seconds. The limit in the surface brightness profiles is
$\mu \sim 27.0 $~mag~arcsec$^{-2}$ (1$\sigma$) using a zero point of
$m_{\rm 0,S^{4}G} = 20.472$.

\begin{figure*}
\includegraphics[width=110mm]{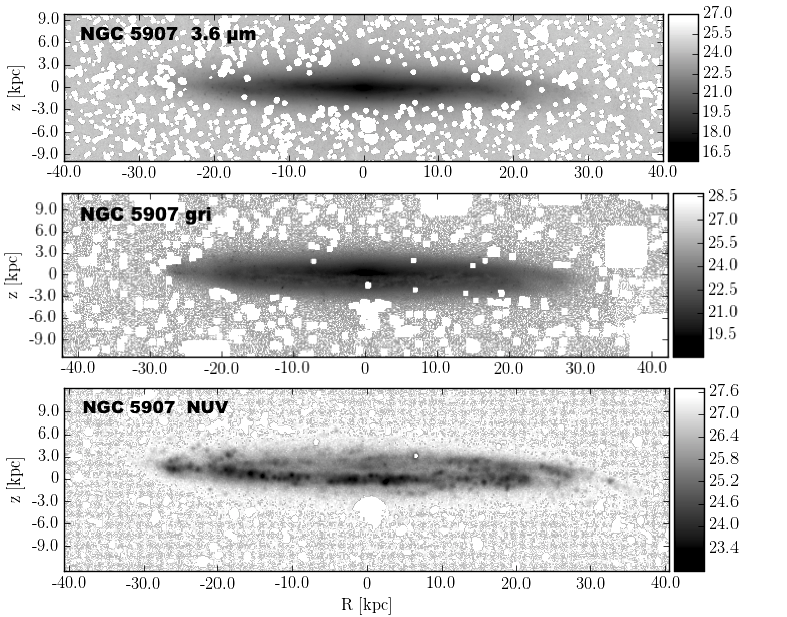}

\caption{Masked images of NGC~5907 in the three wavelength ranges. The greyscale bar is in surface brightness units (mag~arcsec$^{-2}$).}
\label{fig:5907MaskedImages}
\end{figure*}

\section{Methodology}  \label{section:Method}


\subsection{Masking process} \label{subsection:MaskingProcess}

Dealing with deep imaging of the outskirts of galaxies requires careful treatment to remove
the light from foreground and background objects \citep[see e.g.,][]{TrujilloFliri2016}. The goal of the masking
process is to keep only the flux from the sources we want to analyse, avoiding contamination
in their fainter parts. Thus, we mask every fore- and background emission source using a
semi-automatic code (see an example in Fig.~\ref{fig:5907MaskedImages}).

First, using SExtractor \citep{Bertin1996}, the code detects the sources in the
whole image, and builds a mask with all of them. Then, we use Python-based routines to
unmask the galaxy under study. To do that, we convert all the masked areas (except the target) provided from SExtractor to masked pixels. The target is then identified by comparing its coordinates with those of the SExtractor masked regions. Finally, we check visually whether the extent of the mask covers all undesirable flux. For better inspection, we smooth the images and adjust the
contrast to the background limit. This allows us to see if the size of the individual masks
is big enough to cover all light from sources that are not our target. If not, we increase
each masked area in steps of three pixel dilation to completely avoid all the non-desirable flux. By repeating
this process, we are able to remove the background sources from the image. However, some
objects that SExtractor has problems to detect can remain visible around the galaxy under
study, in its outskirts and even within it (e.g., stars from our Galaxy). To get rid of
these, we add masks manually with Python.

This is a common method for all bands, except for the S$^{4}$G data, where the masks were
supplied by the S$^{4}$G team \citep[details in][]{Sheth2010}. In this case, we increase the masked areas using the method described above to cover all foreground and background objects when adjusting the contrast to the background limit. 


\subsection{Sky subtraction and background treatment} \label{subsection:SkySub}

We used background-subtracted images from each of the telescopes. However, once the images were
masked (see Sect.~\ref{subsection:MaskingProcess}), we performed an additional non-aggressive sky
subtraction by measuring the background level as in \citet{PohlenTrujillo2006}. We
determined the gradient of a very extended radial surface brightness profile of each galaxy ($\sim 1.5$ times the size of the galaxy),
convolved with a one-dimensional Gaussian kernel (1 pixel standard deviation). The
mean sky level of each image is the mean value of the most extended part ($\sim 80$~arcsec length) of the radial
surface brightness profile in the corresponding flattest region of its gradient function for
each galaxy, and this is subtracted from the whole image. As we obtain the sky value from
our region of interest through very extended radial surface brightness profiles, this method
allows us to identify clearly where a galaxy ends and, consequently, where a profile
starts to be dominated by the sky background component. Note that our data were free of large scale gradients in the sky background emission, so the procedures of \cite{Peters2017} are not considered here.


\subsection{Profile extraction and truncation position} \label{subsection:SBPextraction}

 \begin{figure*}
\includegraphics[width=\textwidth]{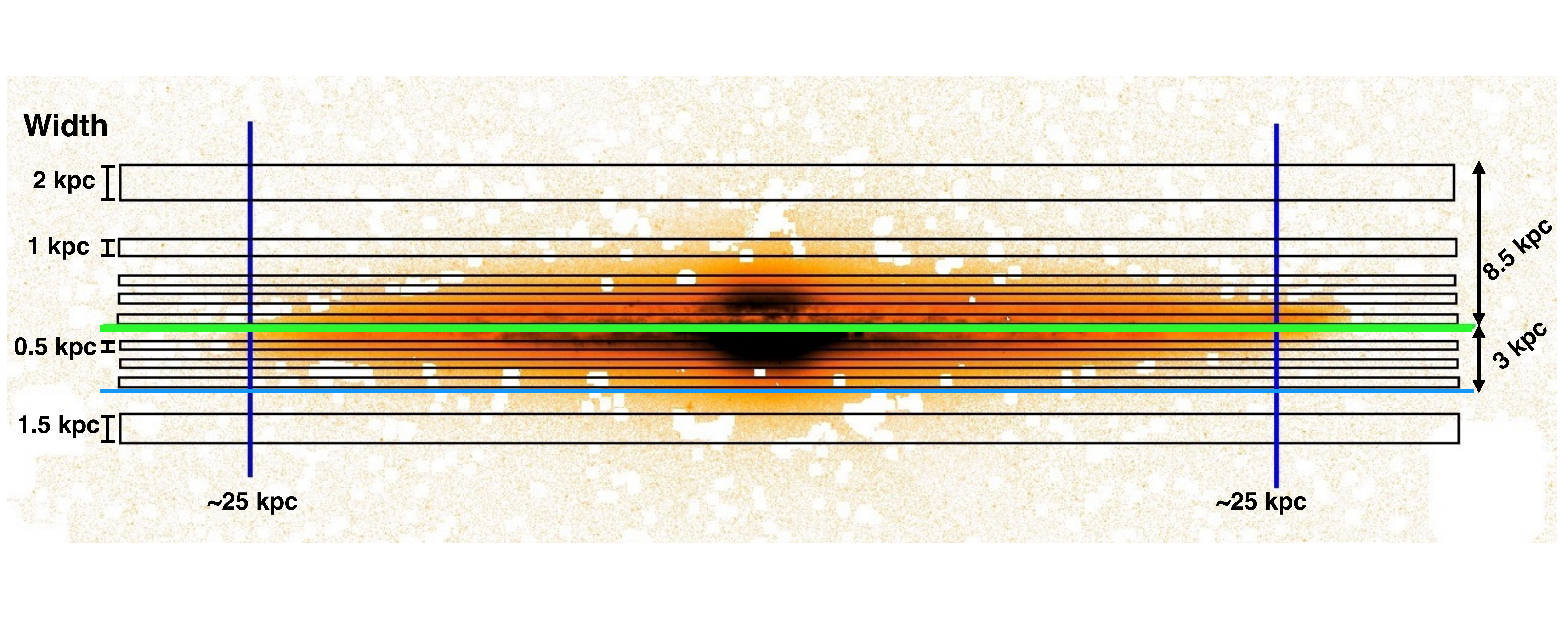}

\caption{Image of NGC~4565 in the SDSS \textit{gri} combined band. The green
horizontal line covers the region along which we obtain the galaxy mid-plane radial surface
brightness profile. The black rectangles show the locations where the radial surface
brightness profiles above and below the galaxy mid-plane were extracted. The numbers on the
left are the widths of each profile. The numbers on the right are (1) the highest altitude we
reach (8.5~kpc) and (2) the height up to where we are able to detect the truncation
(3~kpc) which is indicated with the horizontal blue line. The vertical dark blue
lines point out a radial distance of $\sim 25$~kpc from the centre of the galaxy. The truncation for this galaxy
is located at $\sim 26.4 \pm 0.4$~kpc.}

\label{fig:NGC4565slits}
\end{figure*}

The galaxy surface brightness profiles in the radial direction were
extracted by calculating the flux values along a slit of constant width. The width of the slit was determined in terms of the physical distance
(kpc) in the galaxy, depending on its apparent size and distance. The methodology followed
to obtain the profiles is the same for all the wavelength ranges. The first step was to
rotate the masked and sky-subtracted galaxy image in order to have horizontal major axes. 

To obtain the radial surface brightness profiles, we started with a central
profile in the galactic plane using a slit of 0.5~kpc width (see Fig.
\ref{fig:NGC4565slits}). We calculated the right- and left- band side surface brightness profiles
separately, by dividing the slit into radial logarithmic scale bins, from the centre of the
galaxy to the outer regions. The logarithmic bins allow us to increase the signal-to-noise
ratio (S/N) in the outskirts. For each bin, we calculated the surface brightness correcting
by $A_{\lambda}$, the Galactic extinction coefficient at each wavelength, presented as the
total absorption in magnitudes in the NED. The final surface brightness profile
was obtained by averaging the left and right radial logarithmic bins at the same distance
from the centre of the galaxy. Despite small asymmetries in the radial direction, we combine
the two sides to explore the surface brightness profiles down to very faint levels (below
$\mu \sim 27.0$~mag~arcsec$^{-2}$).

 After producing the central radial profile, we obtained the \textit{shifted} radial surface
brightness profiles at a given height above and below the galaxy mid-plane. We proceed in a
similar manner to the mid-plane profile, but averaging the right and left radial
logarithmic bins of the profile above and below the galaxy mid-plane. The first
\textit{shifted} radial surface brightness profile is obtained at $|z| =0.5$~kpc above and
below the galaxy mid-plane with a height (vertical box) also equal to 0.5~kpc. Then, by
repeating the same procedure we go up/down in steps in height of 0.5~kpc up to where we
start to lose S/N. We consider low S/N when the mean size of the error bars is larger than
the intrinsic oscillations in the profile due to the galaxy structure (i.e., spiral arms,
regions of star formation, etc.), and we recover the S/N by increasing the vertical box
height. As our two galaxies are different in terms of apparent size and distance, the steps
in height above (and below) the mid-plane of the galaxy were as follows (see also Fig.
\ref{fig:NGC4565slits}): for NGC~4565, boxes with 0.5~kpc height up to 3~kpc,
and then steps of 1.0, 1.5 and 2.0~kpc at 4.0, 5.5 and 7.5~kpc altitude, respectively; in
the case of NGC~5907, 0.5~kpc boxes up to $|z|=2.5$~kpc, followed by boxes of 0.7, 1.5 and 2.5~kpc at altitudes of 3.2, 4.7 and 6.7~kpc above the mid-plane, respectively. As we are combining boxes that are above and below the galaxy
mid-plane we explored whether there are any significant differences between these two
regions of the galaxy (see Appendix B). We do not find any
significant difference between the two sides of the galaxies except slightly different surface brightnesses (typically $\Delta \mu_{gri} \sim 0.3$~mag~arcsec$^{-2}$ for NGC~4565 and $\Delta \mu_{gri} \sim 0.1$~mag~arcsec$^{-2}$ for NGC~5907), expected from the fact that the galaxies are not perfectly edge-on (particularly NGC~5907).


The uncertainties in each bin of the profiles were for all cases defined as

\begin{equation}
\sigma_{\rm {bin}}^{2} = \sigma_{\rm {Poiss, bin}}^{2} + \sigma_{\rm {sky, bin}}^{2}	\,\, ,
\end{equation}

\noindent
which is the quadratic sum of the Poissonian error of the bin
($\mathrm{\sigma_{Poiss, bin}}$) and its associated sky error value ($\sigma_{\rm {sky,
bin}}$). The first is the contribution of the Poissonian error in the determination of the
signal in each bin of the profile  

\begin{equation}
\sigma_{\rm {Poiss, bin}}= {\rm RMS}(\bar{I}_{\rm {bin}}-I_{\rm {pix}}) /\sqrt{N_{\rm {bin}}} 	\,\, , 
\end{equation}

\noindent
where $I_{\rm {pix}}$ is the intensity value of a given pixel inside the bin; $\bar{I}_{\rm
{bin}}$ is the mean intensity value of all the pixels in the bin; ${\rm RMS}$ is the root mean square of the expression in brackets; and $N_{\rm {bin}}$ is the
number of pixels inside the bin area. The second source of error is the uncertainty in the
determination of the sky due to fluctuations at the bin area scale. It can be written as

\begin{equation}
\sigma_{\rm {sky, bin}} = {{\rm RMS_{sky}}}/\sqrt{N_{\rm {bin}}},
\end{equation}

\noindent
where ${\rm RMS_{sky}}$ is the root mean square across the whole image. This quantity was
determined over three well-distributed and different background regions of $\sim$\,200\,$\times$\,200~arcsec in the masked image. We made a Bayesiann estimate of the RMS per pixel and the confidence in the mean for each
region by using the Python package \verb|scipy.stats.bayes_mvs| \citep{Oliphant2006}. The
uncertainty of the RMS was based on the average confidence limit for the mean.

The last part of this process was to determine the position of the truncation, that we
define as the sharp edge of a highly-inclined galaxy. We consider that the slope changes at
the position of the intersection between the prolongations of two exponential components
in a surface brightness profile. For an unbiased measurement of the truncation radius, we
followed the steps of \citet[][their Sect.~4.2]{PohlenTrujillo2006}. The scale lengths of 
the exponential profiles before and after the truncation for each galaxy and the various wavelengths and heights above/below the galaxies mid-plane are shown in Tables \ref{tabla:h_4565} and \ref{tabla:h_5907}. The derived truncation radii are almost always consistent with those derived by eye. 

\begin{table*}
\centering
\caption{Scale lengths in kiloparsecs of the exponential profile fits on both sides of the truncation for NGC~4565. For each height $\mid z \mid$ and each wavelength range, we give the numerical value of the exponential fit before and after the truncation radius ($h_{\rm bef}$ and $h_{\rm aft}$).} \vspace{4mm}
\label{tabla:h_4565} 
\begin{tabular}{c|cccccc}
\hline
  & \multicolumn{2}{c}{NUV} & \multicolumn{2}{c}{SDSS-$gri$} & \multicolumn{2}{c}{$3.6 \mu$m}  \\
		 $\mid z \mid$ & $h_{\rm bef}$ & $h_{\rm aft}$ & $h_{\rm bef}$  & $h_{\rm aft}$ & $h_{\rm bef}$ & $h_{\rm aft}$\\
		\hline
0.0 & $-14.8 \pm 0.5$ & $ 1.2 \pm 0.2$ & $8.9 \pm 0.3$ & $ 1.7 \pm 0.2$ & $4.7 \pm 0.2$ & $ 1.5 \pm 0.2$ \\
0.5 & $-84.8 \pm 0.5$ & $1.3 \pm 0.2$ & $8.3 \pm 0.2$ & $ 2.0 \pm 0.2$ & $4.8 \pm 0.2$ & $ 2.0 \pm 0.2$  \\
1.0 & $22.5 \pm 0.3$ & $ 2.1 \pm 0.3$ & $7.5 \pm 0.3$ & $3.0 \pm 0.2$ & $5.4 \pm 0.3$ & $ 2.5 \pm 0.2$  \\
1.5 & $19.3 \pm 0.4$ & $ 3.1 \pm 0.2$ & $7.3 \pm 0.2$ & $ 3.8 \pm 0.3$ & $6.0 \pm 0.4$ & $ 3.0 \pm 0.2$ \\
2.0 & $17.8 \pm 0.3$ & $ 4.1 \pm 0.3$ & $7.6 \pm 0.2$ & $4.0 \pm 0.2$ & $6.5 \pm 0.3$ & $ 4.1 \pm 0.3$ \\
2.5 & $54.7 \pm 0.5$ & $ 6.6 \pm 0.3$ & $8.3 \pm 0.4$ & $ 4.8 \pm 0.2$ & $7.2 \pm 0.4$ & $3.3 \pm 0.3$ \\
3.0 & $40.3 \pm 0.5$ & $5.1 \pm 0.4$ & $9.1 \pm 0.4$ & $ 5.0 \pm 0.3$ & $7.3 \pm 0.4$ & $ 3.7 \pm 0.3$ \\
\end{tabular}
\end{table*}

\begin{table*}
\centering
\caption{As in Table~\ref{tabla:h_4565}, but for NGC~5907.  } \vspace{4mm}
\label{tabla:h_5907} 
\begin{tabular}{c|cccccc}
\hline
  & \multicolumn{2}{c}{NUV} & \multicolumn{2}{c}{SDSS-$gri$} & \multicolumn{2}{c}{$3.6 \mu$m}  \\
		 $\mid z \mid$ & $h_{\rm bef}$ & $h_{\rm aft}$ & $h_{\rm bef}$  & $h_{\rm aft}$ & $h_{\rm bef}$ & $h_{\rm aft}$\\
 		\hline
0.0 & $-147.1 \pm 0.5$ & $1.5 \pm 0.1$ & $6.1 \pm 0.2$ & $1.5 \pm 0.2$ & $3.6 \pm 0.2$ & $ 1.8 \pm 0.1$ \\
0.5 & $20.3 \pm 0.5$ & $ 1.3 \pm 0.2$ & $5.8 \pm 0.2$ & $ 1.6 \pm 0.2$ & $3.8 \pm 0.2$ & $ 2.0 \pm 0.2$  \\
1.0 & $6.8 \pm 0.2$ & $ 2.2 \pm 0.2$ & $5.5 \pm 0.2$ & $1.7 \pm 0.2$ & $4.2 \pm 0.3$ & $ 2.2 \pm 0.2$  \\
1.5 & $7.8 \pm 0.2$ & $ 2.1 \pm 0.2$ & $5.7 \pm 0.3$ & $ 2.0 \pm 0.2$ & $5.1 \pm 0.3$ & $ 2.6 \pm 0.2$  \\
2.0 & $11.3 \pm 0.4$ & $ 2.3 \pm 0.2$ & $7.5 \pm 0.3$ & $ 2.6 \pm 0.3$ & $6.3 \pm 0.3$ & $ 3.2 \pm 0.3$  \\
2.5 & $12.6 \pm 0.4$ & $ 3.6 \pm 0.3$ & $9.9 \pm 0.4$ & $ 2.1 \pm 0.2$ & $7.2 \pm 0.4$ & $ 3.6 \pm 0.3$  \\
3.2 & $16.0 \pm 0.5$ & $ 5.6 \pm 0.3$ & $9.5 \pm 0.4$ & $ 2.0 \pm 0.3$ & $8.0 \pm 0.4$ & $ 4.8 \pm 0.4$  \\

\end{tabular}
\end{table*}


\subsection{The effect of the PSF on the surface brightness profiles}  \label{subsection:PSF}

Before starting the analysis of our findings we need to explore whether the effect
of the point spread function (PSF) can alter the location of the truncation. Although this
may appear unlikely along the mid-plane of the galaxies, it is worth exploring whether
any potential truncation feature above the mid-plane could be the result of a PSF effect.
Another issue to explore is whether the addition of extra light  by the PSF in the
outer regions of the galaxies (beyond the mid-plane) can affect the location of this
feature high above the plane.

The PSF effect is stronger in edge-on than in face-on galaxies
\citep[e.g.,][]{deJong2008, Sandin2014,Sandin2015} due to the
departure from circular symmetry in the edge-on configuration \citep[see a
quantification of the effect of the ellipticity on the PSF effect
in][]{Trujillo2001a,Trujillo2001b}. This has been explored in previous works
\citep[e.g.,][]{Zibetti2004a, Sandin2014, TrujilloFliri2016,
Peters2017, Comeron2017} that showed that the properties of stellar
haloes and faint outskirts of galactic discs can be influenced significantly by the wings of
the PSF.

To check how the faint wings of the PSF are affecting our measurements, we used an extended and deep
SDSS \textit{r}-band PSF ($R=13.2$\,arcmin), built by combining very bright (saturated) and fainter
stars \citep[following the procedure of][]{FliriTrujillo2016}. We then used IMFIT
\citep{Erwin2015} to perform a 2D fit of a truncated exponential
disc for each galaxy, producing PSF-convolved and PSF-deconvolved models of the optical data. We made a two-disc fit (thin and
thick disc) with a bulge and a bar component (the latter only for NGC~4565) according to the morphological classification showed in Table~\ref{tabla:paramfisicos} to properly reproduce both galaxies. In both cases, we needed a thin disc truncated at the same
distance as the data and a pure exponential thick disc. As the dust content is important in the optical (although much less than in UV, see Fig.~\ref{fig:5907MaskedImages}), the central part of the images (i.e., the dust lanes) was masked so the models were determined by extrapolating the analytical functions from the disc regions where the dust is not visible. The 2D fit uses a minimization method based on the Levenberg-Marquardt least-squares algorithm \citep{More1978}. Fig.~\ref{fig:GalColourComposition} shows how our galaxy images are affected by removing the PSF, mainly along the minor axis of both galaxies, where the scattered light is redistributed towards the inner parts. 

\begin{figure*}
\includegraphics[width=170mm]{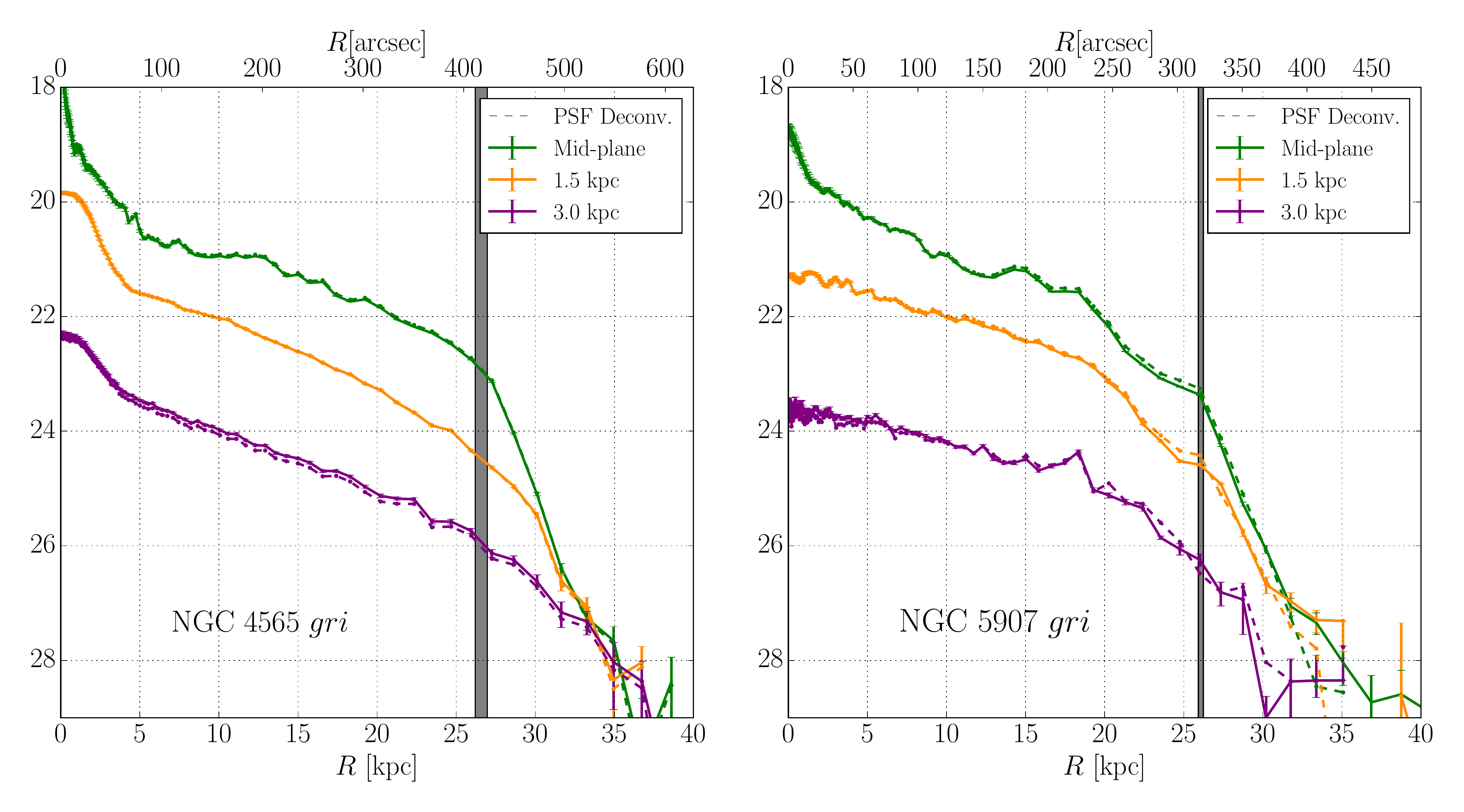}\hspace{6mm}

\caption{Radial surface brightness profiles for each galaxy at different heights above/below
the galaxy mid-plane in the SDSS \textit{gri} combined band for NGC~4565 (left) and NGC~5907 (right). In both panels, the solid curves
are radial surface brightness profiles obtained from the observed data along the galaxy
mid-plane (green), at a height of 1.5~kpc (yellow) and at a height of 3.0~kpc (violet).
The dashed curves represent the same radial surface brightness profiles but for the
PSF-deconvolved model of each galaxy at the corresponding location. The vertical boxes of
all the profiles are 0.5~kpc wide. The vertical dark grey region represents the mean
position of the truncation for all the heights in the SDSS \textit{gri} combined band,
plus/minus the standard deviation of that distribution of truncation positions. }

\label{fig:SBP0153kpc}
\end{figure*}

In order to quantify the effect of the PSF on the surface brightness profiles, we extract
the surface brightness profiles from the deconvolved images following exactly the same slit
configurations as in the original dataset. Fig.~\ref{fig:SBP0153kpc} shows that the position
of the truncation is not affected by the PSF: its value is exactly the same as in the original data
(see Sect.~\ref{section:results}). In Fig.~\ref{fig:SBP0153kpc} we also see the truncation up to
$z=3$~kpc, but it is not more prominent than in the data (see details in Figs.
\ref{fig:4565panelModel} and \ref{fig:5907panelModel}). Thus, the truncation position is not
affected by the PSF in terms of the height above the galaxy mid-plane where it is
detectable, nor in terms of its visibility.

We have explicitly checked the effect of the PSF only on the optical bands as there we have the largest and deepest, and thus most reliable, PSF available (dynamical range of $\sim$20~mag and $R=13.2$~arcmin). Furthermore, the depth of the optical data is similar to IR or NUV (the latter is deeper but with poorer S/N) and any PSF effects will therefore not be much more pronounced in other wavelength than in the optical. As the position of the truncation is independent of wavelength (which implies different telescopes and cameras with both different PSFs and spatial resolutions) and we do not see a large PSF effect in the \textit{r}-filter, we assume that the effect of the PSF is also negligible in the other bands. 

To check the possible effect of the PSF in more detail, we considered a truncated infinitely thin disc (i.e., with a vertical scale height equal to zero), which we modelled according to the size scales in our galaxy sample. In this situation, the scattered light thickens the disc but it does so more effectively in the central parts than in the outskirts. As a result, we see that a decline in the surface brightness profile (which resembles a "truncation"-like feature) occurs at increasingly shorter distances from the centre as we move vertically away from the mid-plane. At a vertical distance of 0.5~kpc, the radial position of this "truncation"-like feature is $\sim 2$~kpc closer than in the mid plane (where it occurs at $R \sim 26$~kpc), while at a height of 3~kpc it is located at $R \sim 10$~kpc.

While this simple test confirms what is expected, namely that a thin disk model gets thicker after convolving it with a PSF, it produces a result that is very different from what we observe: that the truncation above the midplane in real galaxies is at the same radial location at all heights. This test thus strengthens our conclusions that the measured surface brightness at the truncation up to 3~kpc is not dominated by the bright emission coming from the galaxy's mid-plane, that this feature is thus not a PSF artifact, and that the location of the truncation is essentially unaffected by scattered light.

\section{Results} \label{section:results}

\subsection{Truncations are more prominent in NUV than in IR} \label{subsection:ResultsSBprofColour}

\begin{figure}
\includegraphics[width=94mm]{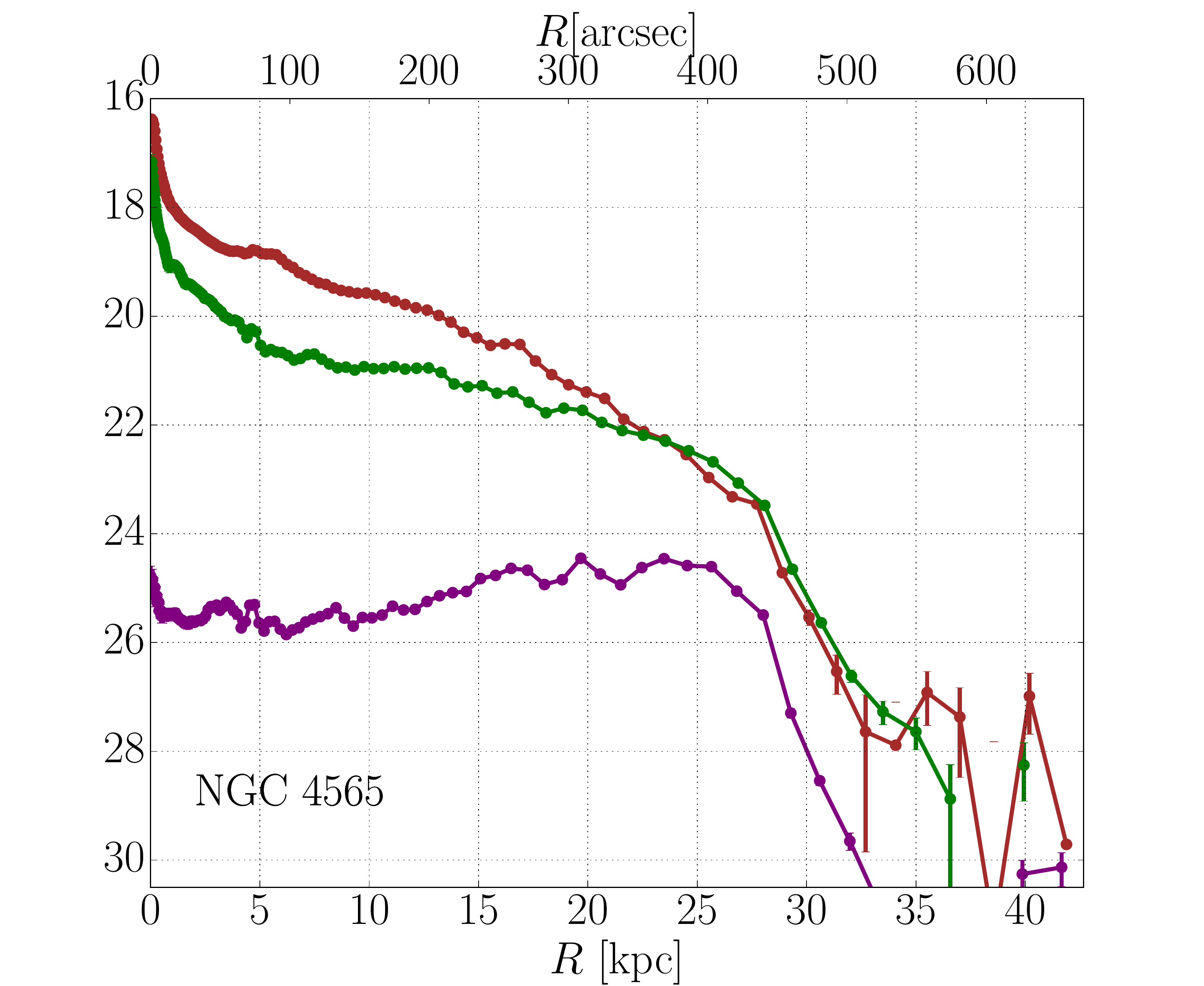}
\includegraphics[width=94mm]{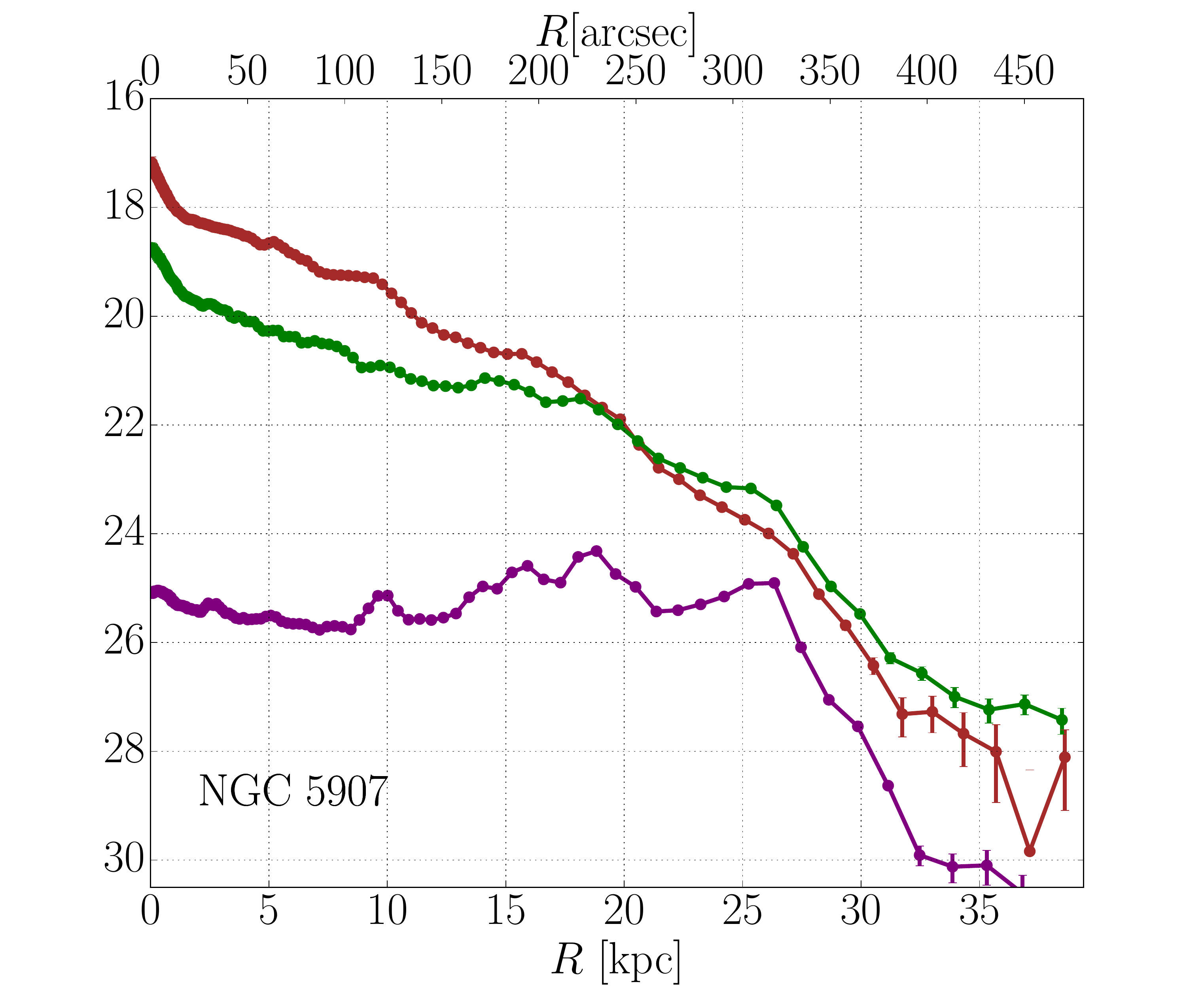}

\caption{Surface brightness profiles along the disc mid-plane of NGC~4565 (up) and
NGC~5907 (down) in the three wavelength ranges studied. In each panel, the three curves correspond, from top to bottom, to NIR (red), optical (green), and UV (purple).}

\label{fig:4565central}
\end{figure}

In Fig.~\ref{fig:4565central} we plot the radial surface
brightness profiles along the mid-plane of NGC~4565 and NGC~5907 in our three wavelength ranges: \textit{GALEX} NUV, SDSS \textit{gri} combined band
and \textit{Spitzer} $3.6 \,  \micron$. The profiles were obtained as described in
Sect.~\ref{subsection:SBPextraction}, where the slit width is 0.5~kpc and the radial bin
size changes logarithmically.   

In both galaxies, a truncation is clearly observed. The change in slope between the inner
region of the disc ($R<R_{\rm{trun}}$) and the outer region ($R>R_{\rm{trun}}$) is more
prominent in bluer bands, and the truncation is most prominent in the NUV band. It is least pronounced,
but still clearly present, in the IR. The main difference in the shape of the surface
brightness profiles is in the flux inside the truncation. Dust in this region
severely affects the NUV and to a much lesser degree, the optical profile, producing a much flatter distribution than
in the NIR regime \citep[see a review of dust extinction in][]{Mathis1990}.

Figure~\ref{fig:4565central} shows that the stellar mass density (traced by the IR band) in both galaxies has a peak
near the centre because of the bulge (more prominent in NGC~4565), decreases
slowly until the truncation, and then falls. If we assume the difference between the
IR and optical profiles to be predominantly due to dust, 
we deduce that the amount of dust along the radial direction is such that (up to 10 kpc) it barely affects the slope of the optical profile (particularly in the case of NGC~5907) compared to the stellar mass density profile provided by the IR band\footnote{See the
approximately constant \textit{gri}-$3.6 \micron$  colour of  $\sim 2$~mag inside 10~kpc in both galaxies in Fig.
\ref{fig:ColourSBP}.}, and then it diminishes further until the position of the truncation. The
similar slope after the truncation for all bands (see Tables \ref{tabla:h_4565} and \ref{tabla:h_5907}) suggests that there is essentially no dust in that regions, and that the star formation rate (SFR) after the
truncation (if anything) is low. Finally, in the NUV band we see that the discs are strongly affected by
dust extinction (see also Fig.~\ref{fig:5907MaskedImages}, lower panel) so the distribution
is mainly flat with some fluctuations due to either star formation activity peaks or less dust extinction up to the truncation position, for both galaxies. Beyond there, the profiles bend over sharply. 

\subsection{Radial SB profiles: truncation position is independent of colour} \label{subsection:ResultsSBprofTruncPos}

\begin{figure}
\includegraphics[width=90mm]{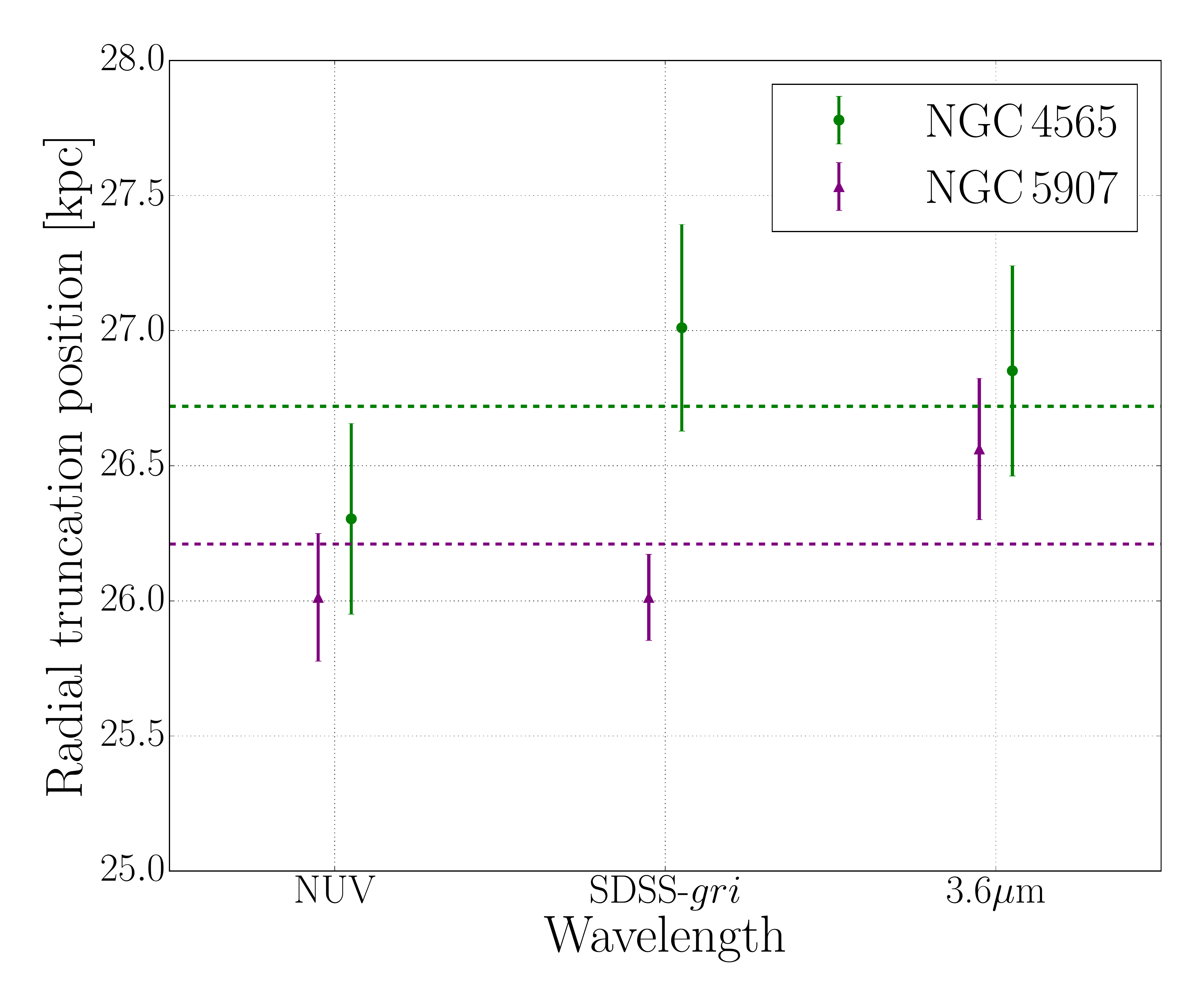}
\includegraphics[width=90mm]{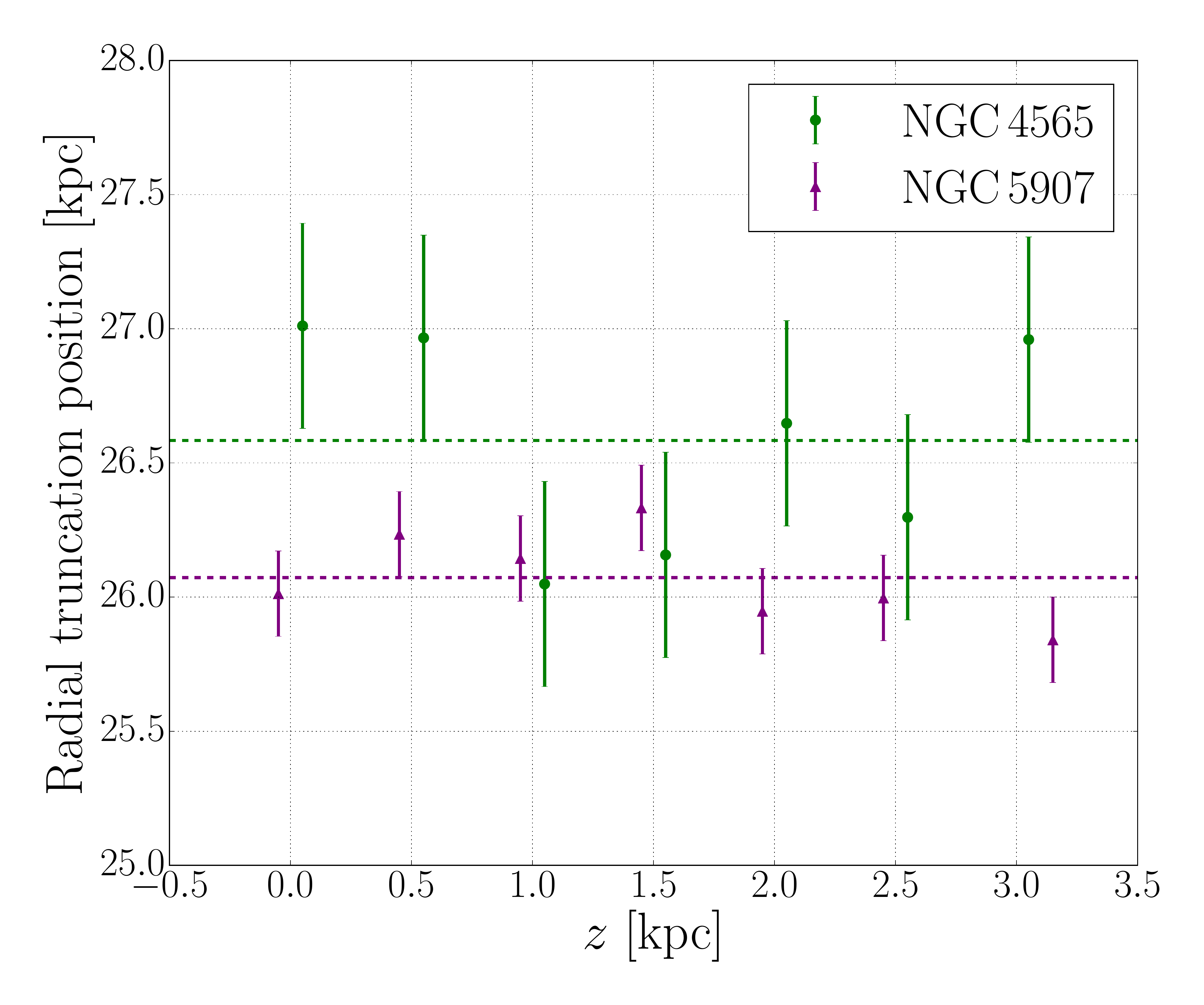}

\caption{Truncation position at each wavelength, measured from the SB profiles along the disc plane (above) and at different heights in the SDSS \textit{gri} combined band (bottom). In both cases the dashed lines represent the mean value of the points in the corresponding colour.}

\label{fig:lambdaVSTruncPos}
\end{figure}

Figure~\ref{fig:lambdaVSTruncPos} shows another noteworthy feature in common between the two
galaxies: the radial location of the truncation is the same (within the uncertainties) in all
three wavelength ranges, NUV (\textit{GALEX}), SDSS \textit{gri} combined band, and
IRAC/\textit{Spitzer} $3.6\,\micron$ (Tables~\ref{tabla:trunPos_4565} and \ref{tabla:trunPos_5907}). We obtain a mean value for the three wavelengths at the galaxies mid-plane of 26.7\,$\pm$\,0.4~kpc for NGC~4565 (26.3$\pm$0.4~kpc for NUV, 27.0$\pm$0.4~kpc in optical, and 26.9$\pm$0.4~kpc in NIR), and of 26.2\,$\pm$\,0.2~kpc for NGC~5907 (26.0$\pm$0.2~kpc in NUV, 26.0$\pm$0.1~kpc in optical and 26.6$\pm$0.2~kpc for the NIR), although these values in physical units are affected by uncertainties in the distance (see Sect.~\ref{section:SelectionData} and Table~\ref{tabla:paramfisicos}). The
independence of the location of the truncation on wavelength suggests that the truncation is a
robust feature of the galaxies. This is because the truncation is observed in profiles as
different as the one that traces the stellar mass density (i.e., 3.6$\mu$m) and the one connected
with star formation activity and dust (i.e., NUV). As we will explore later, this makes the
truncation into an ideal tool both to measure the sizes of the discs and to explore the ongoing growth of the galaxies.

\begin{table}
\centering
\caption{Radial positions of the truncation in kpc at each height $\mid z \mid$, and at each wavelength for NGC~4565. The last column shows the mean values at each height above/below the galaxy mid-plane $\langle R_{\rm trunc}\rangle_{z}$, while the last row shows the average values for each wavelength range $\langle R_{\rm trunc}\rangle_{\lambda}$. $\langle R_{\rm trunc} \rangle_{z, \lambda}$ is the mean truncation position for all $\mid z \mid$ and $\lambda$.} \vspace{4mm}
\label{tabla:trunPos_4565} 
\begin{tabular}{c|ccc|c}
\hline
$\mid z \mid$  & NUV & SDSS-\textit{gri} & $3.6 \micron$ & $\langle R_{\rm trunc}\rangle_{z}$ \\
\hline 
0.0 & $26.3 \pm 0.4$ & $27.0 \pm 0.4$ & $26.9 \pm 0.4$ & $26.7 \pm 0.4$ \\
0.5 & $26.4 \pm 0.3$ & $27.0 \pm 0.3$ & $25.8 \pm 0.3$ & $26.4 \pm 0.3$ \\
1.0 & $25.8 \pm 0.4$ & $26.0 \pm 0.3$ & $25.9 \pm 0.4$ & $25.9 \pm 0.4$ \\
1.5 & $25.9 \pm 0.4$ & $26.2 \pm 0.4$ & $26.1 \pm 0.4$ & $26.1 \pm 0.4$ \\
2.0 & $26.9 \pm 0.4$ & $26.6 \pm 0.4$ & $26.3 \pm 0.5$ & $26.6 \pm 0.4$\\
2.5 & $25.9 \pm 0.5$ & $26.3 \pm 0.4$ & $26.8 \pm 0.5$ & $26.3 \pm 0.5$\\
3.0 & $26.3 \pm 0.5$ & $27.0 \pm 0.4$ & $26.3 \pm 0.4$ & $26.5 \pm 0.4$\\
\hline 
$\langle R_{\rm trunc}\rangle_{\lambda} $ & $26.2 \pm 0.4$ & $26.6 \pm 0.4$ & $26.3 \pm 0.4$ & $\langle R_{\rm trunc}\rangle _{z, \lambda}=26.4 \pm 0.4$ \\
\hline
\end{tabular}
\end{table}

\begin{table}
\centering
\caption{As in Table~\ref{tabla:trunPos_4565}, but for NGC~5907.  } \vspace{4mm}
\label{tabla:trunPos_5907} 
\begin{tabular}{c|ccc|c}
\hline
$\mid z \mid$ & NUV & SDSS-$gri$ & $3.6 \mu$m & $\langle R_{\rm trunc}\rangle_{z}$ \\
\hline 
0.0 & $26.0 \pm 0.2$ & $26.0 \pm 0.1$ & $26.6 \pm 0.2$ & $26.2 \pm 0.2$ \\
0.5 & $25.9 \pm 0.1$ & $26.2 \pm 0.1$ & $26.1 \pm 0.2$ & $26.1 \pm 0.1$ \\
1.0 & $26.7 \pm 0.2$ & $26.1 \pm 0.2$ & $26.0 \pm 0.3$ & $26.3 \pm 0.2$ \\
1.5 & $26.0 \pm 0.2$ & $26.3 \pm 0.1$ & $26.4 \pm 0.2$ & $26.3 \pm 0.2$ \\
2.0 & $25.9 \pm 0.2$ & $25.9 \pm 0.1$ & $25.8 \pm 0.2$ & $25.9 \pm 0.2$ \\
2.5 & $26.0 \pm 0.2$ & $26.0 \pm 0.2$ & $25.9 \pm 0.2$ & $26.0 \pm 0.2$ \\
3.2 & $26.2 \pm 0.3$ & $25.8 \pm 0.2$ & $25.9 \pm 0.3$ & $26.0 \pm 0.3$ \\
\hline 
$\langle R_{\rm trunc}\rangle_{\lambda}$  & $26.1 \pm 0.2$ & $26.1 \pm 0.1$ & $26.1 \pm 0.2$ & $\langle R_{\rm trunc}\rangle_{z, \lambda}=26.1 \pm 0.2$ \\
\hline
\end{tabular}
\end{table}

Interestingly, the surface brightness and the radial position of the truncations are very similar
in both objects (i.e., $\mu_{3.6\mu \rm m} \sim$23~mag arcsec$^-2$ and $R_{ \rm trun}\sim 26.5$~kpc). Although we do not know whether this is just by chance or physically connected to the fact that our two galaxies have the same maximum rotational velocity (i.e., $\sim$230~km s$^-1$).

\subsection{Comparison with previous studies} \label{subsection:ResultsComparSudies}

NGC~4565 and NGC~5907 were the first galaxies in which a truncation was detected and analysed via surface photometry, from photographic plates \citep{vanderKruit1979, vanderKruit1981a, vanderKruit1981b}. They defined the truncation location as a cut-off radius, $R_{{ \rm max}}$, just after a sharp cut-off with an exponential folding of $\lesssim$1~kpc \citep[see Figs.~5 and 11 from][]{vanderKruit1981a}. Their values for $R_{{ \rm max}}$ were 32.6~kpc for NGC~4565 (assumed distance 10~Mpc) and 29.6~kpc for NGC~5907 (for a distance modulus of 11~Mpc). These values were later confirmed with excellent precision using CCD photometry by \cite{Morrison1994} for NGC~5907, and by \cite{Wu2002} for NGC~4565. There is also good agreement with our results if we consider the discrepancies in the truncation (we set the truncation position at the cut-off of the exponential folding), as well as the different distance values.

There is, also good agreement with \cite{Martin-Navarro2012}, who obtained surface brightness profiles for 34
highly inclined spiral galaxies from the S$^{4}$G, including NGC~5907. They find that
NGC~5907 is a type II galaxy (truncated) with its truncation position located at
$R\sim26$~kpc, in agreement with our result taking into account that their assumed
distance to the object (16.3~Mpc) is slightly different from ours (16.9~Mpc). 

\cite{Comeron2012, Comeron2017} studied seventy S$^{4}$G edge-on galaxies, and found, from 1D surface brightness profile fitting of NGC~4565 for the whole disc, but also for each component (thin and thick disc) separately,
that NGC~4565 is a type II+II galaxy, i.e., with two truncations along the disc. The second truncation (in which we are interested) is located at $431 \pm 12$~arcsec ($27.5 \pm 0.8$~kpc) for the whole disc fit. This result again agrees with ours, considering the uncertainties and methodologies in each work.

\subsection{Truncation position is constant above the mid-plane} \label{subsection:ResultsverticalEvolut}

In order to study the vertical behaviour of the position of the truncation, we derived
radial surface brightness profiles at different altitudes above (and below) the galactic
mid-plane in the two galaxies, as explained in Sect.~\ref{subsection:SBPextraction}. In
Appendix \ref{appendix}, we show the set of surface brightness profiles:
Figs.~\ref{fig:4565panelNUV},~\ref{fig:4565panelgri} and~\ref{fig:4565panel36} show the results
for NGC~4565 in the NUV, \textit{gri} and $\mathrm{3.6 \, \micron}$ bands, respectively, while
Figs.~\ref{fig:5907panelNUV},~\ref{fig:5907panelgri} and~\ref{fig:5907panel36} show the
results of NGC~5907. The radial positions of the truncations are shown in Tables~\ref{tabla:trunPos_4565} and \ref{tabla:trunPos_5907} for NGC~4565 and NGC~5907, respectively.

In Fig.~\ref{fig:lambdaVSTruncPos} we summarise this by showing all the positions of the truncation
values along the vertical axis in the case of the SDSS \textit{gri} combined band, for both
galaxies. The position of the truncation in NGC~4565 is practically constant up to a height of
$z\sim3$~kpc. Above that value, we are unable to explore the truncation and the profiles
become compatible with pure exponentials because of the poor S/N. For NGC~5907 we
observe the same behaviour, but up to $z\sim3.2$~kpc. In Appendix \ref{appendix} we show that the position of the truncation is roughly the same for the whole wavelength range, up to those altitudes, in both galaxies. It is important to note that the truncation is always sharpest in the mid-plane and becomes less prominent at larger heights. The average truncation values for all the heights (indicated in Tables~\ref{tabla:trunPos_4565} and \ref{tabla:trunPos_5907} and in Fig.~\ref{fig:lambdaVSTruncPos} with dashed lines for the optical) are: $26.2 \pm 0.4$~kpc (NUV), $26.6 \pm 0.4$~kpc (optical) and $26.3 \pm 0.4$~kpc (NIR) for NGC~4565 and $26.1 \pm 0.2$~kpc (NUV), $26.1 \pm 0.1$~kpc (optical) and $26.1 \pm 0.2$~kpc (NIR) for NGC~5907.

The independence of the location of the truncation with height has been also found in a previous
study: \cite{deJong2007} showed that the truncation (they call it break) occurs at the
same radius independent of the height above the disc up to $z\sim 1.5$~kpc in the low-mass ($v_{\rm{rot}} =98$~km s$^-1$;
$M_{\rm{abs, B}} = -18.4$) edge-on galaxy NGC~4244. This study is based on resolved stellar populations of NGC~4244 by
counting stars in both the F606W and F814W bands of the \textit{HST} Advanced Camera for Surveys
(ACS). They found that the truncation is sharpest in the mid-plane and nearly
disappears at large heights, also in concordance with our results.

In further agreement, \cite{Comeron2012, Comeron2017} found that beyond 3~kpc above
the mid-plane, the surface brightness profiles of NGC~4565 do not show any evidence for
truncation. In particular, they explored the region from $z\sim$42~arcsec
(2.7~kpc) up to $z\sim 110$~arcsec (7.0~kpc) above the galaxy mid-plane.

Our truncation position values are affected by the spatial resolution
of the surface brightness profiles. In NGC~4565, the corresponding surface brightness
profile width is equal to 0.5~kpc. In the case of NGC~5907, as its absolute magnitude is
similar to NGC~4565 but the distance to the object is higher (see
Table~\ref{tabla:paramfisicos}), the S/N is lower at such altitudes. Consequently, the last
profile where we detect the truncation is 0.7~kpc wide. A more conservative
statement would thus be that the location of the truncation is the same at all heights up to
$z \sim 3$~kpc in both galaxies. Above 3~kpc, this feature is not visible, either (1) because it is
where the truncation actually ends and/or (2) the light coming from the thick disc or
from the stellar halo outshines the thin disc.

Finally, we stress again that the location of the truncation as visible up to 3~kpc in both galaxies occurs at a similar position. As previously stated, this may either be a consequence of the similar characteristics of the galaxies or simply a coincidence.\\


\subsection{Radial colour profiles} \label{subsection:ResultsColourSBP}

\begin{figure*}
\includegraphics[width=\textwidth]{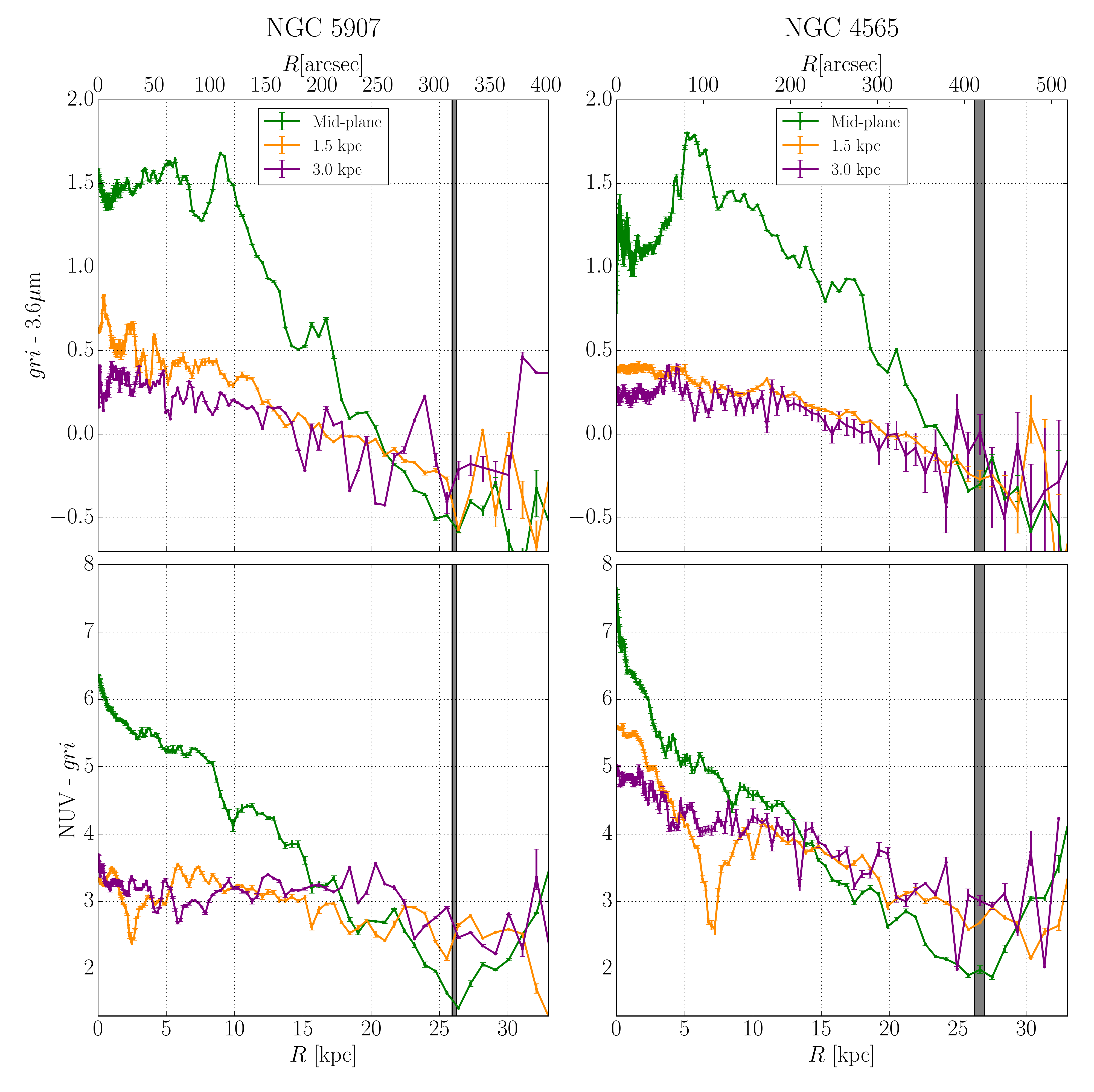}

\caption{Radial colour profiles for both galaxies along their mid-plane, at 1.5~kpc, and at
3~kpc height (NGC~5907 at left and NGC~4565 at right). The rows show the different colours: \textit{gri}\,-\,3.6$\micron$ in the top row
and NUV\,-\,\textit{gri} in the bottom. The vertical dark grey regions of each galaxy
represent the mean position of the truncations for all the heights in the SDSS \textit{gri}
combined band, plus/minus the standard deviation of that distribution of truncation
positions: 26.6$\pm$0.4~kpc for NGC~4565 and 26.1$\pm$0.2~kpc for NGC~5907.}

\label{fig:ColourSBP}
\end{figure*}

From the radial surface brightness profiles in the three wavelength ranges, we can obtain
radial colour profiles. In Fig.~\ref{fig:ColourSBP} we show the \textit{gri}\,-\,3.6$\micron$
and the NUV\,-\,\textit{gri} radial colour profiles for both galaxies along the mid-plane,
at 1.5~kpc and at 3~kpc height above/below it. The colours are similar (though
bluer in the mid-plane) at the truncation radii for the three altitudes in all cases.
Inward of the truncation, the colours show a negative gradient (less clearly for
NUV\,-\,\textit{gri} in NGC~5907) with a very similar behaviour in the radial profiles
at 1.5~kpc and 3~kpc height above/below the mid-plane. The mid-plane profiles are strongly
affected by dust, mainly in the inner parts of both galaxies. Beyond the truncation, in the
NUV\,-\,\textit{gri} mid-plane colour, we can observe an increase in the colour values for the two
galaxies\footnote{There is a hint that the same behaviour could also happen in the
\textit{gri}\,-\,3.6$\micron$ colour, although the trend is not so obvious as in the
NUV\,-\,\textit{gri} colour.}. This U shape around the truncation resembles the behaviour
found at the break position in previous works \citep[see
e.g.,][]{Azzollini2008a,Bakos2008}. We will take advantage of this feature below to
explore the ongoing growth of these galaxies.

\subsection{The vertical extent of the thin disc}

Our analysis of the surface brightness profiles above the galaxies mid-plane shows that it is
possible to identify the truncation up to at least 3~kpc above/below the disc plane. Moreover, the
location of the truncation, within the uncertainty in determining this position, is the same
independently of the altitude above the disc. It is difficult to understand this without
invoking a similar physical origin for the truncation in both the galactic plane and at
higher altitudes. The obvious conclusion is then that the thin disc extends up to at least 3~kpc
in height. Thus, we can use the maximum height where the truncation has been
identified to provide a lower limit to the vertical extent of the thin disc. 

The above result does not imply that there is no need for a thick disc in our two galaxies - our result only suggests that the brightness of any thick disc component is not enough to outshine the truncation feature from the thin disc up to at least 3~kpc height. In this sense, we can make a direct comparison of our results with those of \citet{Comeron2017}, who found that the ratio of thick to thin disc mass is a declining function of the total mass of the galaxy as measured by the circular maximum velocity. In the galaxy we have in common, NGC~4565, \citet{Comeron2017} quote a mass ratio between the two disc components of 14\%. In addition, they provide the surface brightness profiles of both the thin and thick component (see their Appendix). At the location of the truncation, they report $\Sigma_T(0)/\Sigma_t(0)$=0.04, and scale heights $h_t$=7.1~arcsec (0.45~kpc) and $h_T$=49.3~arcsec (3.15~kpc) for the thin and thick discs, respectively. If we assume an exponential decay for both disc components in the vertical direction: $\Sigma_t(z)=\Sigma_t(0)\times\exp(-z/h_t)$ (thin disc) and $\Sigma_T(z)=\Sigma_T(0)\times\exp(-z/h_T)$ (thick disc), these numbers imply that the thick disc starts to dominate the light of the thin disc at $z = 1.7$~kpc. However, making reasonable assumptions on the dilution of the thin disc truncation by thick disc light (e.g., the thin disc 10 times fainter than the thick disc), we can estimate that the truncation can be observed up to $z \sim 2.9$~kpc, in very nice agreement with our observed result that we detect the truncation of the thin disc at altitudes as high as 3~kpc. In future work, we will use deeper data (with better S/N) of NGC~4565 to investigate whether the truncation can be detected at even larger distances above the mid-plane.

\section{The ongoing growth of galactic discs} \label{section:growth}

In this Section we will discuss two methods to derive a consistent measure of the ongoing growth of galactic discs. In the first, we will use the location of the truncation at different heights above the mid-plane. The second method analyses the shape of the radial colour profiles beyond the truncation.

\subsection{First approach: offsets in the location of the truncation at different heights}  \label{section:firstApp}

\begin{figure}
\includegraphics[width=85mm]{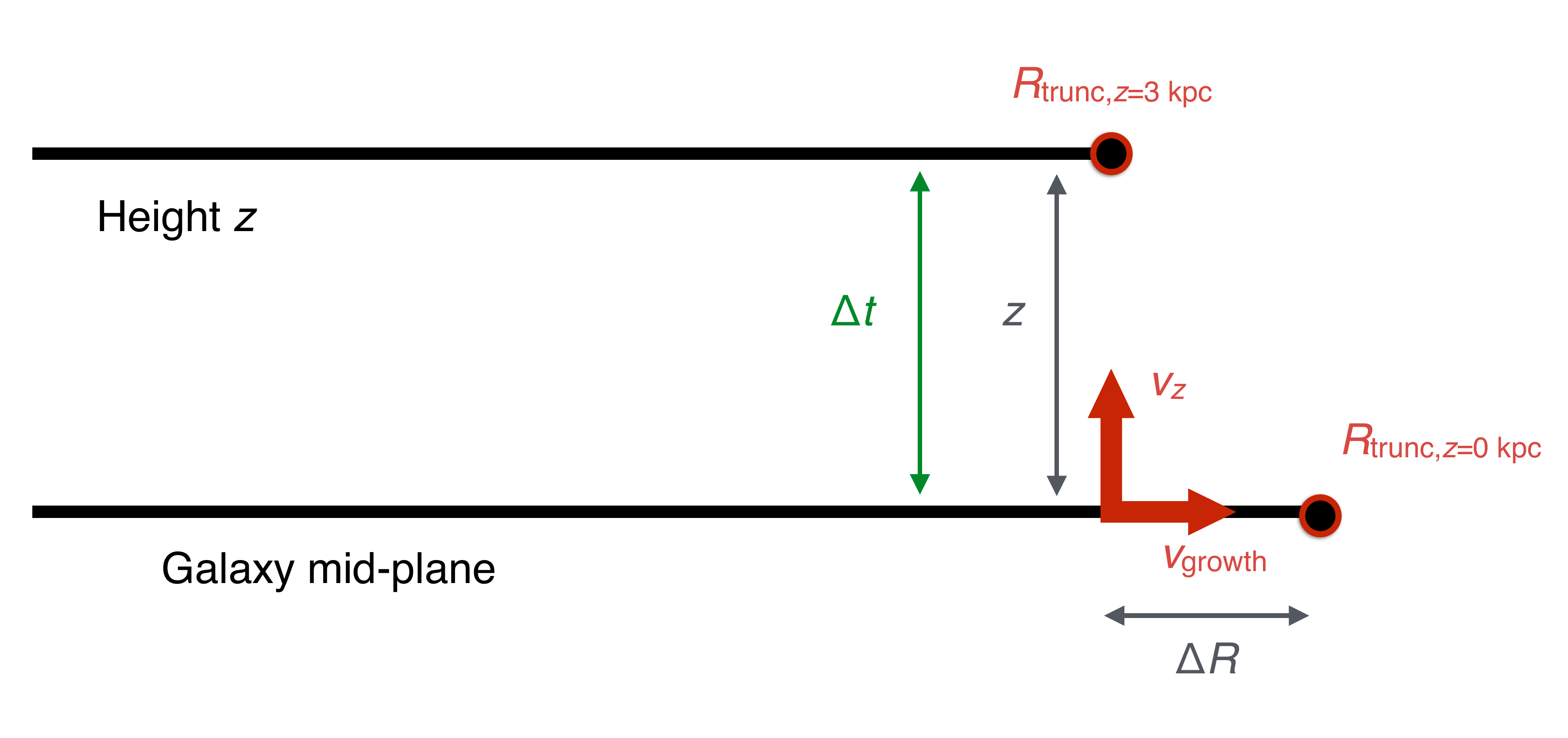}

\caption{A cartoon illustrating how to use the different vertical locations of the radial position of
the truncation to measure the ongoing growth of the galaxies. Stars take a time $\Delta t$ to
reach a certain height $z$ above the galaxy mid-plane. During such time, the location of the
truncation in the mid-plane moves a radial distance $\Delta R$. In this particular example,
we use $z$=3~kpc to illustrate this idea.}

\label{fig:verticaltruncationoffsets}
\end{figure}

Stars are subject to migrations both in the radial and vertical directions \citep[see for a review][]{Sellwood2014, Debattista2017}. Stars currently being formed in the mid-plane at the truncation position will take some time to reach a given height
above the mid-plane of the galaxy. For this reason, considering an inside-out propagation of the star
formation, the location of the truncation at a high location should be closer
to the galactic centre than that in the mid-plane. We illustrate this with a cartoon in
Fig.~\ref{fig:verticaltruncationoffsets}.

We call $v_{\rm growth}$ the speed with which the truncation moves radially
outwards in the mid-plane. Over an amount of time $\Delta t$, the position
of the truncation in the mid-plane moves a radial distance $\Delta R$. As stars born at the
truncation in the mid-plane take a time to reach a given height $z$, we can use the
different position of the truncation at height $z$ compared to the mid-plane to measure the speed
of galaxy growth. The key element is then to know how much time it takes a typical star to
reach a certain height $z$ at a given radial distance $R$. Once this is known, $v_{\rm growth}$ will simply be

\begin{equation}
v_{\rm growth}=\frac{\Delta R}{\Delta t} \, .
\end{equation}

\noindent
In order to estimate $v_{\rm growth}$, we estimate how long it takes for a given
star in the disc mid-plane to reach a height of 3~kpc (i.e., we calculate $\Delta t$). We
use $z= 3$~kpc as this is approximately the maximum height at which we are able to measure
the truncation above the mid-plane for both galaxies. To estimate $\Delta t$, we calculate
the difference in gravitational potential between a star located at a radial position $R$ and
$z$=0, and a star at the same radial distance $R$ and $z$=3~kpc. This way, we can
estimate the initial vertical velocity $v_{z}$ that a star needs to reach a height $z$
starting from the mid-plane $v_{z}$.

The gravitational potential for an infinitely thin disc is given by
\citet[][]{Toomre1963}

\begin{equation}
\Phi (R, z) = -2 \pi G \int dk \: {\rm exp}(-k |z|) J_{0}(kR) S(k) \, ,
\end{equation}

\noindent
where $S(k)= \int dR \Sigma (R) R J_{0}(kR)$ is the Hankel transform of $\Sigma (R)$, and
$J_{0}(kR)$ is the zero-order first-kind Bessel function. 

We consider an exponential disc with surface density 

\begin{equation}
\Sigma (R) = \dfrac{\alpha ^{2} M}{2 \pi}  e^{-\alpha R} \, ,
\end{equation}

\noindent
where $M$ is the total disc mass and $\alpha$ is the inverse of the disc scale length ($h$).
Thus, if we call $\Sigma_{0}=\alpha ^{2} M/2 \pi$, the resulting gravitational potential as
a function of $R$ and $z$ is

\begin{equation}
\Phi (R, z) =  \dfrac{-2 \pi G \Sigma_{0}}{\alpha ^{2} }   \int_{0}^{\infty } dk  \:
\dfrac{J_{0}(kR) e^{-k |z|}}{[1+k^{2} \alpha ^{-2}]^{3/2}} \, . 
\end{equation}

\noindent
As our galaxies are similar to the Milky Way, we use for a rough estimate of $v_{z}$
the structural properties of the disc of our Galaxy. In other words, we replace the galaxy's physical
parameters by those of the Milky Way, from \cite{BlandHawthorn2016}

\begin{equation}
\ M=3.6 \times 10^{10} M_{\sun} \quad ; \quad h=2.6 \,{\rm kpc} \, .
\end{equation}

\noindent
We solve numerically the integral of the gravitational potential and we then assume that the corresponding potential energy at $R$=26~kpc and $z$=3~kpc is equal to the kinetic energy at the same radial distance but at $z$=0~kpc. Hence, taking the values above, we find $v_{z}=8.1$~kpc\,Gyr$^{-1}$ (i.e., $v_{z}\sim 7.9$\,km\,s$^{-1}$) at $R$=26~kpc.

Once we have the gravitational potential distribution along the disc, the gravitational
acceleration at any ($R$, $z$) position is given by 

\begin{equation}
\ g(R, z) = - \nabla \Phi (R, z) \, .
\end{equation}

\noindent
Having $v_{z}$ we are now ready to estimate $\Delta t$. We divide the trajectory of the star from $z$=0 to $z$=3~kpc in steps of 0.1~kpc. This simplifies
our measurement, as we can assume constant acceleration in such small vertical steps and we
can use the motion equations for a particle under uniform acceleration. With the above $v_{z}$,
the amount of time a star at $R=26$~kpc needs to reach $z=3$~kpc is $t\sim0.4$\,Gyr. 

These give us now the possibility to estimate $v_{\rm growth}$. In reality, as we are
unable to measure $\Delta R$ with our current uncertainties, we use those uncertainties to
provide an upper limit to the maximum growth speed of the discs in our two galaxies. For
NGC~4565 we use $\Delta R <$0.25~kpc and for NGC~5907 $\Delta R <$0.4~kpc. With such
values, the maximum speed for the growth of our galaxies is: $v_{\rm growth} <$1~kpc
Gyr$^{-1}$ for NGC~4565 and $v_{\rm growth} <$0.6~kpc~Gyr$^{-1}$ for NGC~5907.

We stress the approximations that we have used in our
estimate of $v_{\rm growth}$. The most relevant one is that we have neglected the effect of the
gravitational potential of both the bulge component and the dark matter halo of these
galaxies. This is justified as we have only explored the difference in gravitational
potential in a region very close to the galactic disc (i.e., $z$=3~kpc) and at a large distance
from the center ($R$=26~kpc). At this location, the difference in radial distance between a star
in the mid-plane and one at $z$=3~kpc is just 0.2~kpc and the gravitational
potential of the bulge is basically the same in both locations. Similarly, assuming a spherical dark matter
halo, the small difference in the radial position with respect to the galactic centre
implies that the variation of the gravitational potential for such structure is small. Another source of uncertainty in our estimate is the use of the stellar mass and radial scale length of the MW to model these two galaxies. We base this approximation on the
fact that their rotational velocities are similar to the one in our own Galaxy.

We now discuss how reasonable the value of $v_{z}$ is that we have estimated. To
do this we use the fact that we know the surface brightness at the location of the
truncation, both at $z$=0 and $z$=3~kpc; at $z$=0: $\mu_{3.6 \mu \rm m} \sim$23~mag arcsec$^-2$, and at $z$=3~kpc: $\mu_{3.6 \mu \rm m} \sim$26~mag arcsec$^-2$. Consequently, the difference in stellar mass density, assuming a similar $M/L$ ratio
for both locations, is a factor of 16. In other words, stars are 16 times more
abundant in the disc mid-plane than at 3~kpc height at the location of the truncation. We assume now that the vertical velocity distribution of stars at the location of the
truncation is Gaussian. Under this hypothesis, the vertical stellar distribution would be also well approximated by a Gaussian distribution with a width similar to the vertical velocity distribution one. As the density of stars at 3 kpc is a factor of 16 smaller than at $z$=0~kpc, assuming that the stellar vertical density declines as a gaussian distribution 3 kpc will then correspond to a height of $\sim$2.4$\sigma$ over the plane. Now, it is straightforward to estimate the vertical velocity distribution  $\sigma_{z}$ knowing that at 3 kpc, the aforementioned value of $v_{z}\sim 7.6$\,km\,s$^{-1}$ corresponds to 2.4$\sigma_{z}$. Consequently,  $\sigma_{z}$$\sim$3.4~km~s$^{-1}$. Unfortunately, there
are no measurements of the vertical velocity dispersion of the stars at radial distances as
large as $R$=26~kpc. We can only compare this number with the typical velocity dispersion of
the thin disc at the radial location of the Sun ($R$=8.2$\pm$0.1~kpc):
$\sigma_{z}^{t}$=25$\pm$5~km~s$^{-1}$ \citep[][]{BlandHawthorn2016}. As expected, the estimated velocity dispersion at $R$=26~kpc is significantly smaller than at $R$=8~kpc. As the velocity dispersion is expected to decrease as the square root of the mass density, the mass density at the Solar position should then be $\sim$(25/3.4)$^2 \sim$54 times larger than at the truncation. As the
local mass surface density at the Solar position is 70$\pm$5~$M_{\odot}$~pc$^{-2}$ \citep[][]{BlandHawthorn2016}, our prediction is that the local mass surface density at the truncation should be $\sim$1.3~$M_{\odot}$~pc$^{-2}$.  We can check whether such value is consistent with the stellar mass densities one can derive from the stellar population properties at the truncation position.

We have used two different methods for estimating the stellar mass density based on the stellar population properties. The first one uses the stellar population models in \citet{Vazdekis2016}, assuming a bimodal initial mass function of slope 1.3 (similar to the one of Kroupa, \citealt{Kroupa2001}) and Solar metallicity. We extract the $M/L$ ratio at $\mu_{3.6\mu \rm m}$ for the age obtained from the NUV-\textit{gri} colour profile (note that the $M/L$ at $\mu_{3.6\mu \rm m}$ is almost independent of metallicity; see Figs.~13 and 14 in \citealt{Roeck2015}) for each galaxy. From this and the surface brightness at $3.6\mu \rm m$ corrected for inclination \citep[which is around 2~mag;][]{Martin-Navarro2014} we obtain the stellar mas density at the position of the truncation. Alternatively, we can also use the \textit{g}-\textit{r} colours as well as the surface brightness in $r$-band (corrected for inclination) at the truncation and derive a M/L using the prescription by \cite{Bell2003}. Using both methods we estimate the stellar mass density at the truncation position to be within the following ranges: 1.4-2.4~$M_{\odot}$~pc$^{-2}$ for NGC~4565, and 0.4-1.4~$M_{\odot}$~pc$^{-2}$ for NGC~5907. These numbers are consistent with a local mass surface density at the truncation of $\sim$1.3~$M_{\odot}$~pc$^{-2}$, even though we are using simple approximations.

\subsection{Second approach: the shape of the radial colour profile beyond the truncation}

A completely independent way to measure the growth speed of galaxies is by exploring the shape of
the radial colour profile beyond the truncation. The main assumption that we make in this
approach is that there is negligible new star formation beyond the truncation. This
seems to be a very good approximation, considering the sharp decline in all the surface
brightness profiles, particularly in the NUV, after the truncation. If our hypothesis is correct then we can
interpret the gradual reddening colour of the stars beyond the truncation as purely driven by
the passive aging of their stellar populations. We also assume that beyond the truncation
the dust obscuration is not playing any role, in agreement with the idea that there is no
star formation beyond this point.

\begin{figure}
\includegraphics[width=90mm]{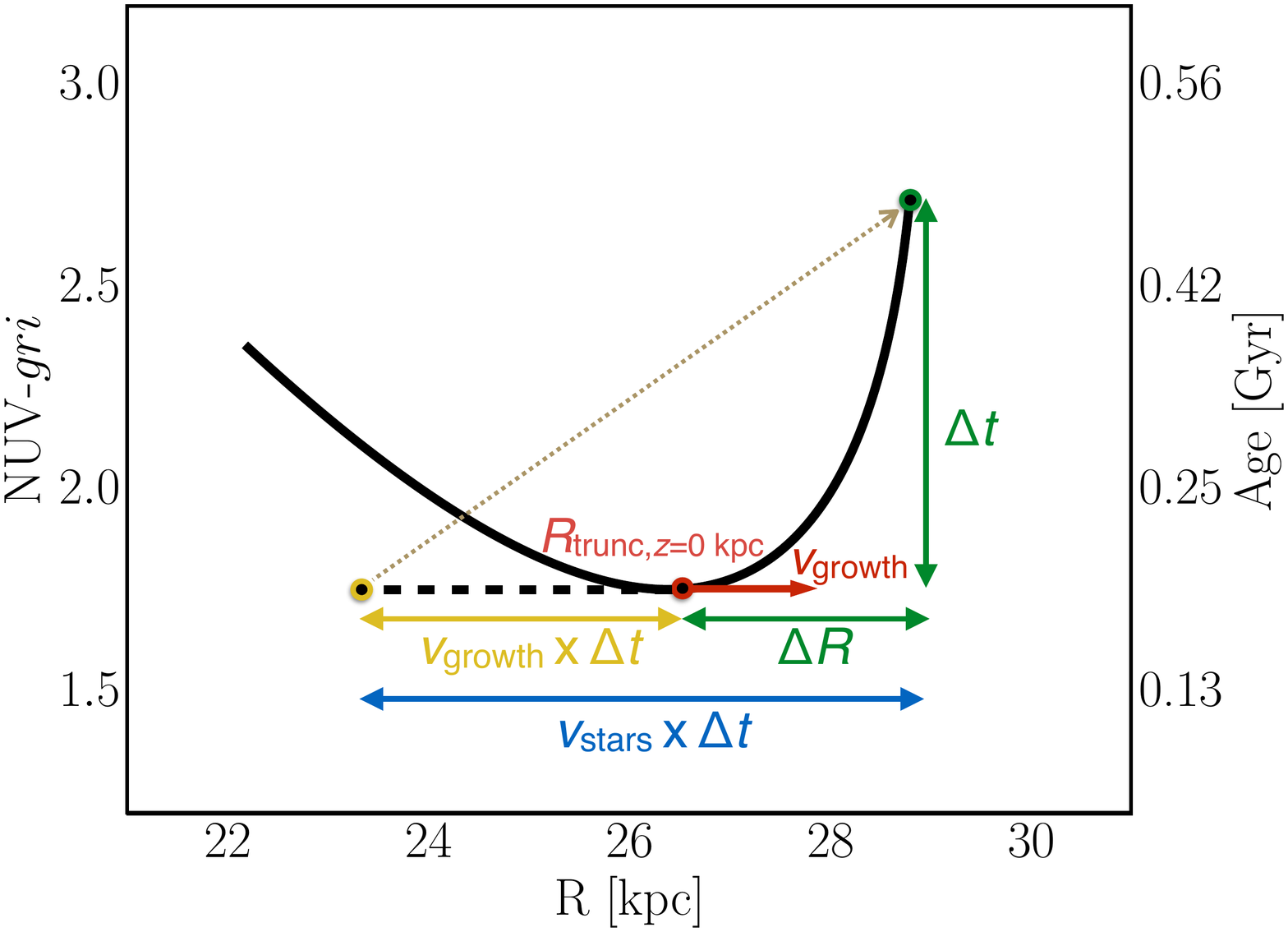}

\caption{A cartoon illustrating how to use the shape of the radial colour profile beyond the
truncation to estimate the growth speed $v_{\rm growth}$ of a disc. The observables in the
radial colour profile beyond the truncation are $\Delta t$ and $\Delta R$. In this particular
example we use the colour NUV-\textit{r} (vertical left axis) as a proxy to measure the interval of time
(vertical right axis). The yellow dot indicates the birth location of a star in the colour-radial distance plane. The green dot is the current position of the star born at the yellow
dot location. The red dot is the position of the truncation.}

\label{fig:colorspeed}
\end{figure}

\begin{figure}
\includegraphics[width=90mm]{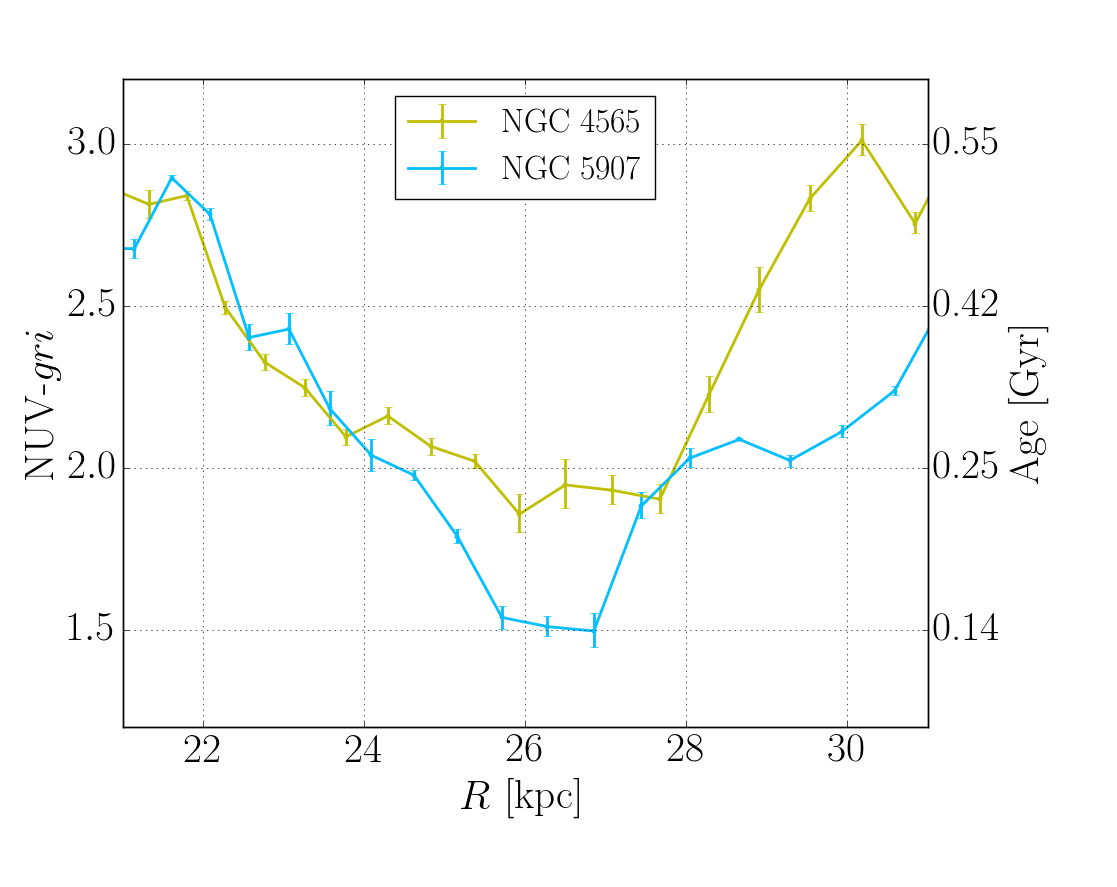}

\caption{A zoom in to the location of the truncation in the NUV-\textit{gri} colour versus radial location plane. The right vertical axis shows the age corresponding to the NUV-\textit{r} colour based on the stellar populations models from \citet{Vazdekis2016}.}

\label{fig:truncationzoom}
\end{figure}

In order to estimate the growth speed of the disc $v_{\rm growth}$ we follow the scheme presented
in Fig.~\ref{fig:colorspeed} (based on the data from Fig.~\ref{fig:truncationzoom}). We use the NUV-\textit{r} radial colour profile\footnote{Actually, we use the
colour profile NUV-\textit{gri} to gain S/N in the outer regions of the galaxies. We have confirmed
that the combined \textit{gri} surface brightness profile is equivalent to the \textit{r}-band to a level below 0.1~mag
along the entire profile of the galaxies.} as a proxy for measuring the time interval
$\Delta t$. We use this colour as this combination of bands is very sensitive to recent star formation and consequently allows us to measure intervals of time of the order of 100~Myr. In our scheme, stars are born at a radial position $R$ (which coincides with
the initial location of the truncation at the beginning of the time interval). After a given
time $\Delta t$, the stars have moved outwards a radial distance
$v_{\rm stars} \times \Delta t$. At the same time, the truncation has also moved outwards by $v_{\rm growth} \times \Delta t$. It is then straightforward to show that

\begin{equation}
v_{\rm growth}=v_{\rm stars}-\frac{\Delta R}{\Delta t} \, .
\end{equation}

\noindent
The values $\Delta R$ and $\Delta t$ can be inferred directly from the data. The
interval $\Delta t$ is measured using $\Delta$(NUV-\textit{gri}). In Fig.~\ref{fig:truncationzoom}
we highlight the location of the truncation region for our two galaxies.

We use the following measurements of colour and radial distance to derive $\Delta R$ and
$\Delta t$. For the galaxy NGC~4565: (27.6$\pm$0.2~kpc, 1.9$\pm$0.1~mag) and (28.3$\pm$0.2~kpc, 2.2$\pm$0.2~mag); and for the
galaxy NGC~5907: (26.9$\pm$0.1~kpc, 1.5$\pm$0.2~mag) and (27.4$\pm$0.1~kpc, 1.9$\pm$0.2~mag), which corresponds to the intervals: $\Delta R$=0.6$\pm$0.2~kpc and $\Delta t$=0.1$\pm$0.01~Gyr for NGC~4565; and $\Delta R$=0.6$\pm$0.1~kpc and $\Delta t$=0.09$\pm$0.01~Gyr for NGC~5907. The transformation from NUV-\textit{gri} colour to age has been done using the \citet{Vazdekis2016} models, assuming Solar metallicity and a bimodal initial mass function (as in Sect.~\ref{section:firstApp}). Consequently, for both galaxies we have  $\Delta R/ \Delta t \sim$6$\pm$2~kpc~Gyr$^{-1}(i.e. \sim$6$\pm$2~km~s$^{-1}$).

As a next step, we need to calculate $v_{\rm stars}$. Unfortunately, there are no measurements
of this value at such locations in galaxies like ours. We can use the ratio of velocity
dispersions at the Solar location, $\sigma_R^t / \sigma_z^t$=35/25 \citep{BlandHawthorn2016},
and assume that this ratio is the same at the location of the truncation. This assumption is quite uncertain, but we can use such a value to illustrate the method. If the above assumption is reasonable, and taking $\sigma_z^t \backsimeq 3.4$~km~s$^{-1}$ from Section~\ref{section:firstApp}, then the typical velocity dispersion at the radial position of the truncation would be $\sigma_R^t $=7/5$\times \sigma_z^t$=7/5$\times$3.4~km~s$^{-1} \sim 5$~km~s$^{-1} \sim v_{\rm stars}$. This value is very similar to the quantity $\Delta R / \Delta t$ we have estimated above, indicating that $v_{\rm growth}$ should be small
(i.e., $v_{\rm growth}<$0.5~kpc~Gyr$^{-1}$). We acknowledge that this estimate is strongly
dependent on the exact value of $v_{\rm stars}$ used, that is unfortunately not known.

Compared to the approach shown in the previous Section, the colour-based method to measure
$v_{\rm growth}$ is much more uncertain and depends strongly on the assumptions we have made.
For example, our measurement of $\Delta t$ is based on the inference of age associated to
the colour NUV-\textit{gri} which depends on the properties of the stellar population model used. As a
way to test the robustness of measuring $\Delta t$ from the colour NUV-\textit{gri}, we can compare the
predictions for such an interval of time based on a completely different stellar population
model \citep{Han2007} that accounts for the binary population. According to
this model, NUV-\textit{r}=1.5~mag corresponds to 0.14~Gyr, NUV-\textit{r}=1.9~mag to 0.21~Gyr and NUV-\textit{r}=2.2
mag to 0.28~Gyr, implying $\Delta t$=0.07~Gyr for NGC~4565 and $\Delta t$=0.07~Gyr
for NGC~5907. These values are similar to the ones inferred using the \citet{Vazdekis2016}
model (i.e., $\sim$0.1~Gyr). We conclude that, as we have stated before, the weakest point in
this approach is the estimate of $v_{\rm stars}$.

In the Introduction we reported the possibility of a common origin for breaks and truncations by considering the predictions in \cite{Rovskar2008a}. Their simulations suggest that the U shape in the colour profiles is the result of stellar migration as the stars can gain angular momentum and move outwards, or lose it and move inwards. This is a stochastic movement, both inwards and outwards, where the average displacement scales with $t^{1/2}$ for a stellar population with a given age. Hence, only the oldest stars get to the larger radii which means that the colours become redder when going outwards, but the break radius may not necessarily be growing. In this sense, our growth rate results might be degenerate as the interpretation of a growing disc is not unique. However, it is important to highlight that in \cite{Rovskar2008a} the breaks are located at radial distances more than two times smaller than our results. We find the truncation at $R_{\rm trunc} \sim 26$~kpc, while the simulations of \cite{Rovskar2008a} report $R_{\rm break} \sim 10$~kpc . The corresponding stellar mas density at the position of the break, $R_{\rm break} \sim 10$~kpc, is $\sim 10$~$M_{\odot}$~pc$^{-2}$ (see Fig.~1 in \citealt{Rovskar2008a}), while here the stellar mass densities at the truncation position are $1.4-2.4$~$M_{\odot}$~pc$^{-2}$ (NGC~4565) and $0.4-1.4$~$M_{\odot}$~pc$^{-2}$ (NGC~5907). The observational work on type II breaks of \citet[][see their Fig.~1]{Bakos2008} is in very good agreement with the predictions of \cite{Rovskar2008a}, which suggests that it is worth exploring whether the features that we are tracing in this work are the same as these in the simulations by \cite{Rovskar2008a}.

\begin{figure}
\includegraphics[width=90mm]{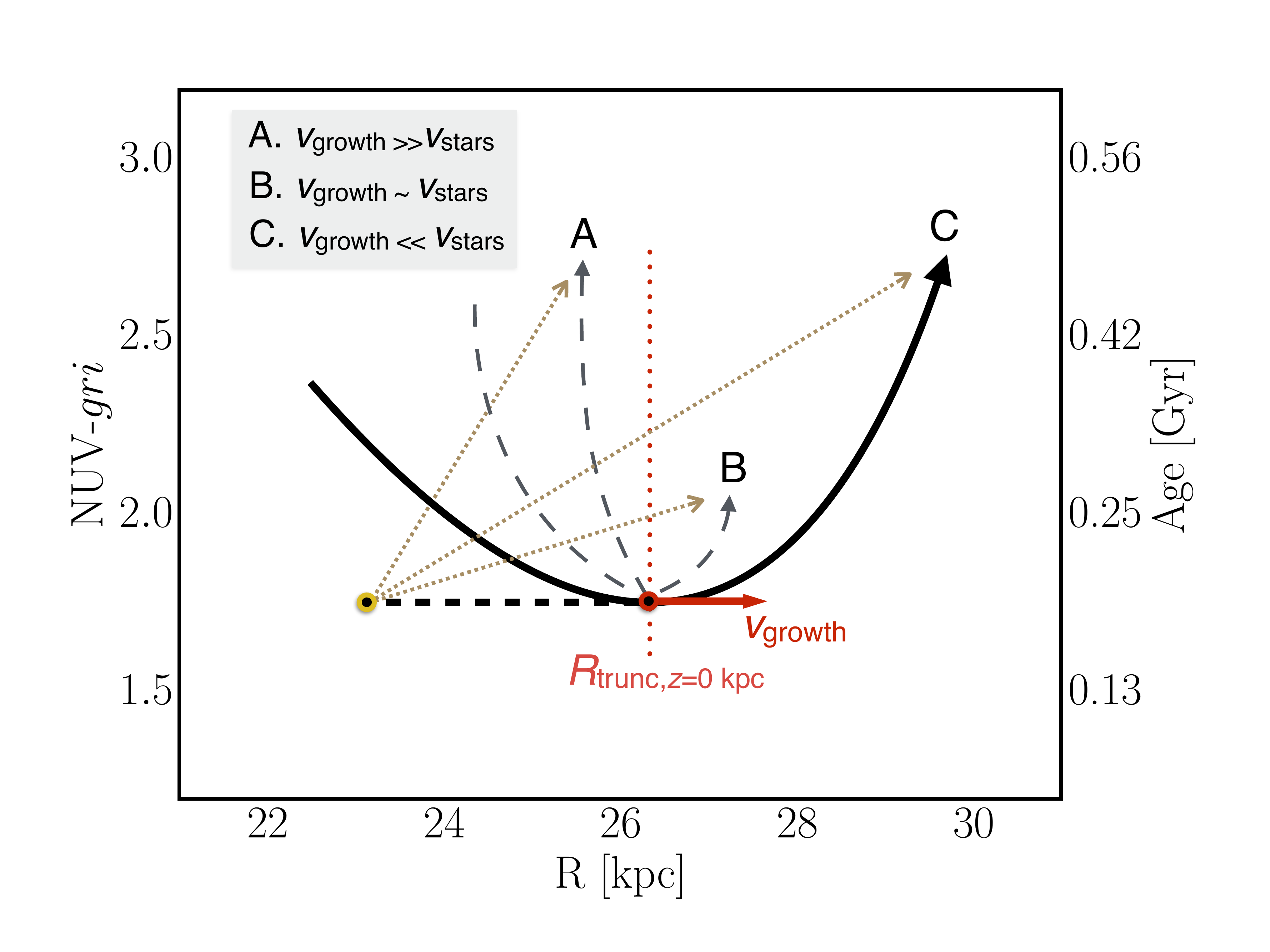}
\caption{A cartoon illustrating the shape of the radial colour profile beyond the truncation in terms of the disc growth speed $v_{\rm growth}$ and the stellar migration speed $v_{\rm stars}$. The yellow dot indicates the birth location of a star in the colour-radial distance plane. The red dot is the current position of the truncation. The three dark yellow arrows (square dotted lines) represent the paths of the new-born star, depending on the growth and migration speeds. The star will end up either in position A, B, or C if its migration speed is smaller than, similar to, or greater than the disc growth speed, respectively.}
\label{fig:GrowthMigrat}
\end{figure}

In Fig.~\ref{fig:GrowthMigrat} we show qualitatively how the shape of the radial colour profile would be in the vicinity of the truncation in terms of the disc growth speed $v_{\rm growth}$ and the stellar migration speed $v_{\rm stars}$. We see that the radial colour profile beyond the truncation becomes steeper when the disc growth speed is larger than the migration speed of the stars. Otherwise, the slope is smoother. By measuring the slope of the radial colour profile after the position of the truncation, we can thus set an upper limit to the growth speed. Comparing Fig.~\ref{fig:GrowthMigrat} with Fig.~\ref{fig:truncationzoom}, we see that the slope in the colour profiles of our data (Fig.~\ref{fig:truncationzoom}) is more similar to the case in that the disc growth speed $v_{\rm growth}$ is lower than the migration speed of the stars $v_{\rm stars}$ (case C in Fig.~\ref{fig:GrowthMigrat}), in agreement with our results.

\section{A comparison with the cosmological growth of discs}  \label{section:comologicalGrowth}

Although it would be extremely interesting to measure, to the best of our knowledge no one has
probed how the radial location of the truncation is moving with cosmic time. The most closely related approach was by \citet{TrujilloPohlen2005}, later expanded by \citet{Azzollini2008}. In those works, the type II breaks of galaxies were
incorrectly called truncations. Nowadays, we know that the type II breaks correspond to a
feature in galaxy discs at closer radial distances than the truncations \citep[see a
full discussion on this issue in][]{Martin-Navarro2012}. While acknowledging that the origin of
the breaks and the truncation could be different, we can nevertheless use the growth of the discs
inferred from the movement of the position of the type II breaks to put our current
measurements of the speed of the truncation into a context.

According to Fig.~8 in \citet{Azzollini2008}, present-day MW-like galaxies
(M$_{\star}\sim$5$\times$10$^{10}$~$M_{\odot}$) would have their break at
$R_{\rm break}\sim$11~kpc. At redshift $\sim$1, objects with the same stellar mass would typically
have $R_{\rm break}\sim$8~kpc. However, a galaxy like the MW had a lower stellar mass 8~Gyr ago.
This means that we can expect $R_{\rm break}$ to be smaller than 8~kpc for the progenitor of the MW
at redshift $\sim$1. Consequently, the 3~kpc increase in the location of the position of the break in
the last 8~Gyr must be considered a lower limit for their speed. In other words, the average
cosmic speed for the location of the break for MW-like galaxies was $>$0.4~kpc~Gyr$^{-1}$, and
probably not larger than 0.5~kpc~Gyr$^{-1}$ assuming that the progenitor of the MW at
redshift $\sim$1 was two times less massive than it is now. In this sense, if the current growth speed of
the truncation were comparable with the cosmic growth in the location of the break, we would
not expect velocities larger than 0.5~kpc~Gyr$^{-1}$. This number is in nice agreement with
our upper limits for the speed of the truncation found above.

\section{What is the physical origin of the truncation?}  \label{section:physicalOrig}

\subsection{Maximum angular momentum or star formation threshold?} \label{subsect:physicalOrigAngMom}

Multiple ideas have been discussed in the literature to understand the origin the truncation
of galaxy discs. The two main concepts have been that the position of the truncation reflects the maximum angular momentum of the protogalactic cloud at the moment of collapse \citep{VanDerKruit1987} or, alternatively, that the location of the
truncation reflects the location of the disc where there is a threshold in the star
formation activity \citep[][]{Kennicutt1989}. In both scenarios, the presence of stars beyond the truncation can
be understood if there is a mechanism of radial migration \citep[see for a review][]{Sellwood2014, Debattista2017} at work. A direct consequence of radial migration would be that, even if originally the angular momentum distribution of the stars were reflecting that of the protogalaxy, a substantial redistribution of the angular
momentum is expected to take place due to secular evolution processes. For example, bars may
play an important role in such a redistribution, as suggested by \cite{Debattista2006}
and \cite{Erwin2008}. Additionally, transient spiral arms cause greater angular
momentum changes to stars at co-rotation than to those at the Lindblad resonances
\citep{Sellwood2002} and can thus move co-rotating stars radially without heating the disc.
\citet{Rovskar2008b} showed that this mechanism is the one responsible for the largest
amounts of redistribution because it is very efficient and can operate on relatively short
timescales. 

The analysis of the properties of truncations that we have made in this paper can shine some
light on the physical processes that are taking place at such locations in galaxies. The
first thing to note is that we have detected for the first time a U shaped colour at the
location of the truncation. This feature \citep[found previously at the
location of breaks in Type II galaxies,][]{Azzollini2008a,Bakos2008} is
considered to be associated to a star formation threshold followed by a migration of stars
to the outer regions \citep[][]{Rovskar2008b, Rovskar2008a}. It thus seems that the truncation in the two MW-like galaxies we have analysed is  closely related to a star formation threshold. It is worth noting that the surface brightness at the
position of the truncation is significantly lower than the typical brightness of the breaks in Type II galaxies. In fact, the average surface brightness of the break in the \textit{r}-band is 22~mag arcsec$^-2$ for galaxies with the brightness of our objects \citep[see Fig.
9 in][]{PohlenTrujillo2006} whereas here we find for the truncation (after deprojecting the
brightness by the effect of the inclination) 25.5~mag arcsec$^-2$. This corresponds to a
factor of $\sim$25 less in stellar mass density for the truncation with respect to the
break. Using the NUV-\textit{gri} and \textit{g}-\textit{r} colours at the location of the truncation we get a stellar
mass density of $\sim 1-2$~$M_{\odot}$~pc$^{-2}$. This stellar mass density is at the bottom
edge for the critical gas density value assumed to trigger the star formation \citep[i.e., 
$\sim 3-10 M_{\odot}$~pc$^{-2}$;][]{Schaye2004}, which in turn implies that the star
formation at this location is quite efficient ($\sim$30\%). 

Both the stellar mass density at the location of the truncation and the U shaped colour profile at
this position suggest that the truncation is linked to a star formation threshold. The
presence, however, of the truncation up to a height of $z \sim 3$~kpc, where the density is
much lower than in the mid-plane, seems to be difficult to explain as the result of a
threshold of star formation at such an altitude. In fact, the characteristic U shape is not
observed at high altitudes (see Fig.~\ref{fig:ColourSBP}). This indicates that truncations
seen at high altitudes may rather be the result of migration of stars from the mid-plane to
higher altitudes.

\subsection{Dust absorption} \label{subsect:physicalDust}

Dust absorption along the mid-plane, especially in the NUV and optical bands, might be evaluated as a potential reason for the presence of a truncation as the dust might cause a flatter apparent gradient in the galaxies' mid-plane radial surface brightness profiles than the true stellar surface density gradient. In theory, this could cause the truncation feature to be just an illusion, created by the largest radius with 
substantial dust content in the mid-plane. The radial colour profiles in Fig.~\ref{fig:ColourSBP} show the clear effect of the dust absorption in the galaxies' mid-planes, mainly in the inner parts. However, the radial colour profiles of NGC\,5907 at the mid-plane, at 1.5~kpc, and at 3~kpc converge on a radius of $\sim 15-18$~kpc and have similar colours until the truncation position, where the U shape appears in the mid-plane colour profiles. Thus, the dust-free regions (i.e., at high altitudes above/below the galaxy mid-plane) have the same colours than the mid-plane from $\sim 15-18$~kpc in this particular case. For NGC\,4565, we find a similar behaviour but from $\sim12-15$~kpc for the NUV\,-\,\textit{gri} colour profiles and from $\sim23$~kpc for \textit{gri}\,-\,3.6$\micron$. Furthermore, we find that the radial position of the truncation is similar across all wavelengths and up to 3~kpc height above/below the mid-plane of both galaxies (see Sects.~\ref{subsection:ResultsSBprofTruncPos} and \ref{subsection:ResultsverticalEvolut}). All these results point to the truncation being a real feature, unaffected by dust absorption.

\subsection{Warps} \label{subsect:physicalWarps}

Another possible origin for the truncations (in the mid-plane) is the presence of warps in the discs of galaxies \citep[e.g.,][]{vanderKruit1979, Naeslund1997}, as one might expect a truncation-like signature at the area where the disc warp begins. Both of the galaxies in our sample have a warp so a proper analysis is necessary. \cite{Rupen1991} reports a warp in the \textsc{HI} emission starting at about 7 arcmin distance from the centre of the galaxy on both sides of NGC~4565. An optical warp that follows the \textsc{HI} warp has been discussed by \cite{vanderKruit1979} and \cite{Naeslund1997}. A warp in the \textsc{HI} gas and in the optical in NGC~5907 has also been reported in several works \citep[e.g.,][]{Shang1998}, starting at 5 arcmin radius \citep[see also][]{Florido1992}. Fig.~\ref{fig:GalColourComposition} shows how the disc isophotes are bending away from the mid-plane near the end of the bright disc component, suggesting a warp-like shape. In the case of NGC~4565 the warp is more clear in the upper right part of the image although it can be seen in the left side of the galaxy disc towards lower regions (Q2 and Q3, respectively in Fig.~\ref{fig:4565_Q}). For NGC~5907 the warp seems to occurs in the opposite way, from the upper left part of the disc towards the lower right region (Q1 and Q4, respectively in Fig.~\ref{fig:5907_Q}). Our observations thus confirm the presence of optical warps at the position of the truncations in the mid-plane.

A potential concern is whether the truncations in the mid-plane could be hidden while observing the galaxy in face-on projection. In order to study this possibility, we have explored whether the position of the truncation continues to be at the same radial location when we explore the galaxy regions with and without the warp. Thus, in Appendix \ref{appendixC} (Fig.~\ref{fig:sbp_warp}), we present an example of the surface brightness profiles at two different heights above/below the mid-plane (0.5 and 2~kpc) in the three wavelength ranges (NUV, \textit{gri} combined band, and 3.6$\micron$), through the warped and non-warped regions, for both galaxies (indicated in Figs.~\ref{fig:4565_Q} and \ref{fig:5907_Q}, for NGC~4565 and NGC~5907, respectively). In these figures, we can see that the presence of a warp barely affects the surface brightness profiles of the galaxies. The only major change is observed in the galaxy NGC~4565 in the NUV profile. There, the presence of a warp pushes the position of the truncation at 2~kpc height towards a larger radial distance compared to the surface brightness profiles closer to the mid-plane. This effect is not seen in the other galaxy, nor in the redder wavelengths.

The results presented here suggest that the warp is populated by very young stars. In the case of NGC~4565 this is confirmed by the stellar population analysis conducted by \cite{Radburn-Smith2014} where they found the warp stellar population to be $<$600~Myr old. In this sense, if one were observing the galaxies at moderate inclinations using optical or infrared bands, the truncations will still be clearly observed at the same radial location than in the edge-on case. We conclude that the presence of a warp does not affect our major conclusion that the location of the truncation at different heights is basically at the same position as in the galaxies' mid-plane. In an upcoming paper (C. Mart\'\i nez-Lombilla et al., in prep.) we will use new ultra-deep imaging of NGC~4565 to address the optical properties of its warp in more detail and relate them to the observed truncation.

\subsection{Stars beyond the truncation} \label{subsect:physicalStarsBeyTr}

We can now consider the number of stars beyond the truncation in the
mid-plane of the galaxies. Using the mid-plane 3.6$\mu \rm m$ surface brightness profile (a good
proxy for the stellar mass density that is barely affected by the dust) we find fractions of 0.2\%
(NGC~4565) and 0.1\% (NGC~5907). We can make a crude estimate to see whether such an amount of
stars is compatible with the idea of radial migration from the region of the truncation to
the outer part of the disc. As we have said before, we expect a radial velocity for the
stars at the radial position of the trucation of $\sim 6 \pm 2$~kpc~Gyr$^{-1}$. In practical terms
this means that the stars produced at the location of the truncation can fill very quickly the outer regions of the galaxies (beyond the truncation). Considering $v_{\rm growth} < 0.6$~kpc~Gyr$^{-1}$ and a typical stellar mass density of $\sim$~1~$M_{\odot}$~pc$^{-2}$, the total mass of stars in an annulus around the truncation (i.e., inner radius $R_{i}=26$~kpc and outer radius $R_{o}= 26.5$~kpc) is $\sim 0.8 \times 10 ^{8}$~$M_{\odot}$. This represents, assuming the total stellar mass of a typical MW galaxy of $5 \times 10 ^{10}$~$M_{\odot}$, a fraction of $\sim 0.15$\%. These crude numbers show that there
are enough stars at the location of the truncation to fill the outer parts of the discs, simply by the process of migration.

\subsection{Type II breaks} \label{subsect:physicalBreaks}

Finally, we consider whether a star formation threshold origin for the truncation
may contradict the presence of a profile break in galaxies, that has also been related to 
star formation activity. One of the most important results by
\citet[][]{Martin-Navarro2012} is that, for galaxies with rotational velocities above
$v_c> 100$~km~s$^{-1}$ (see their Fig.~13), the ratio between the location of the truncation
and the break is quite constant, at around $1.8$. This could indicate that both features
could be linked with different resonances like transient spiral arms \citep{Rovskar2008b}. The fact that both
features correlate strongly with the rotational velocities of galaxies reinforces this view.\\

To summarize, our observational results, together with other evidence from the
literature, points to a scenario where both the origin and the location of the truncation is directly linked to a
threshold in the star formation activity. This threshold is due to the gas density at such a
location which ultimately is connected to the angular velocity of rotation at that radial distance in the galaxy. Migration of stars from their birthplace leads both to the truncation observed well above and below the disc mid-plane, and to the small but significant quantity of stars observed beyond the truncation radius.

\section{Conclusions}  \label{section:conclusions}

We have explored the nature of the disc truncations in two edge-on nearby MW-like galaxies:
NGC~4565 and NGC~5907. We have used \textit{GALEX}, SDSS and \textit{Spitzer} S$^{4}$G imaging to extract radial surface brightness
profiles in and above the mid-planes of the galaxies. We have found the following results:

\begin{itemize}

\item Truncations are observed at the same spatial location ($\sim 26 \pm 0.5$~kpc within the
spatial uncertainties) at all wavelength ranges (NUV, optical and 3.6~$\micron$) in both galaxies.

\item The truncations are observed at the same spatial location (within the
uncertainties) up to a vertical location above the mid-planes of 3~kpc.

\item At the location of the truncation, the NUV-\textit{gri} colour radial profile shows the
characteristic U shape associated with a star formation threshold, in combination with the migration of
stars to the outer regions of the galaxies.

\item The stellar mass density at the location of the truncation is $\sim 1-2~M_{\odot}$~pc$^{-2}$. This number is very close to the critical gas density necessary to transform gas into stars of $\sim 3-10~ M_{\odot}$~pc$^{-2}$, indicating an efficiency of $\sim 30$\% for this
process at the radial location of the truncation.

\item The number of stars in the mid-plane beyond the truncation amounts to $0.1-0.2$\% of the total
stellar mass density of these galaxies. Migration of stars formed inside the truncation radius to the outer
regions can explain this number.

\item The vertical extent of the truncation as well as the colour properties in the
mid-plane beyond the truncation are consistent with an upper limit for the current growth speed for the discs of
the galaxies of $0.6-1$~kpc~Gyr$^{-1}$.

\end{itemize}

\section*{Acknowledgements}

We thank Javier Rom{\'a}n and Ra{\'u}l Infante for providing the extended SDSS
\textit{r}-band PSF, and Sebasti{\'e}n Comer{\'o}n for a number of useful suggestions. We
acknowledge support from the Spanish Ministry of Economy and Competitiveness (MINECO)
under grant numbers AYA2013-41243-P, AYA2016-76219-P and AYA2016-77237-C3-1-P. J.H.K. and I.T.C. acknowledge
financial support from the European Union's Horizon 2020 research and innovation programme
under Marie Sk$\l$odowska-Curie grant agreement No. 721463 to the SUNDIAL ITN network and from the Fundaci{\'o}n BBVA under its 2017 programme of assistance to scientific research groups, for the project ``Using machine-learning techniques to drag galaxies from the noise in deep imaging". J.H.K. thanks the Leverhulme Foundation for the award of a Visiting Professorship at Liverpool John Moores University.

Funding for SDSS-III has been provided by the Alfred P. Sloan Foundation, the Participating
Institutions, the National Science Foundation, and the U.S. Department of Energy Office of
Science. SDSS-III is managed by the Astrophysical Research Consortium for the Participating
Institutions of the SDSS-III Collaboration including the University of Arizona, the
Brazilian Participation Group, Brookhaven National Laboratory, University of Cambridge,
Carnegie Mellon University, University of Florida, the French Participation Group, the
German Participation Group, Harvard University, the Instituto de Astrofisica de Canarias,
the Michigan State/Notre Dame/JINA Participation Group, Johns Hopkins University, Lawrence
Berkeley National Laboratory, Max Planck Institute for Astrophysics, New Mexico State
University, New York University, Ohio State University, Pennsylvania State University,
University of Portsmouth, Princeton University, the Spanish Participation Group, University
of Tokyo, University of Utah, Vanderbilt University, University of Virginia, University of
Washington, and Yale University.

This work is based on archival data from the \textit{Spitzer Space Telescope} available in
the NASA/ IPAC Infrared Science Archive, which is operated by the Jet Propulsion Laboratory,
California Institute of Technology, under contract with the National Aeronautics and Space
Administration. Some of the data presented in this paper were obtained from the Mikulski Archive for Space
Telescopes (MAST). STScI is operated by the Association of Universities for Research in
Astronomy, Inc., under NASA contract NAS5-26555. Support for MAST for non-HST data is
provided by the NASA Office of Space Science via grant NNX09AF08G and by other grants and
contracts.This research has made use of different Python packages and of NASA's Astrophysics Data
(NED) System Bibliographic Services which is operated by the Jet Propulsion Laboratory,
California Institute of Technology, under contract with the National Aeronautic and Space
Administration. 

We acknowledge constructive remarks by an anonymous referee.
 



\bibliographystyle{mnras}
\bibliography{All_references_papers.bib} 



\appendix
\newpage
\section{Surface brightness profiles} \label{appendix}

In the following figures, we plot the radial surface brightness profiles for each galaxy, in each wavelength range and at different heights above/below the galaxy mid-plane. We plot all the surface brightness profiles of a galaxy, in a given band, together, so as to highlight the variations of the truncation position along the vertical axis. Then, we show the same kind of plot but for a PSF-deconvolved model of each galaxy in the optical SDSS \textit{gri} combined band.

In all panels of each figure, the black line shows the mid-plane surface brightness profile. The vertical dark grey region always represents the mean position of the truncation for all the heights in the wavelength range, plus/minus the standard deviation of that distribution of truncation positions. From top to bottom and from left to right, the panels show in red the surface brightness profile at a given altitude above the galaxy mid-plane (indicated in the legend) and, in light grey, the surface brightness profiles plotted in the previous panels.

\begin{figure*}
\centering
\includegraphics[width=\textwidth]{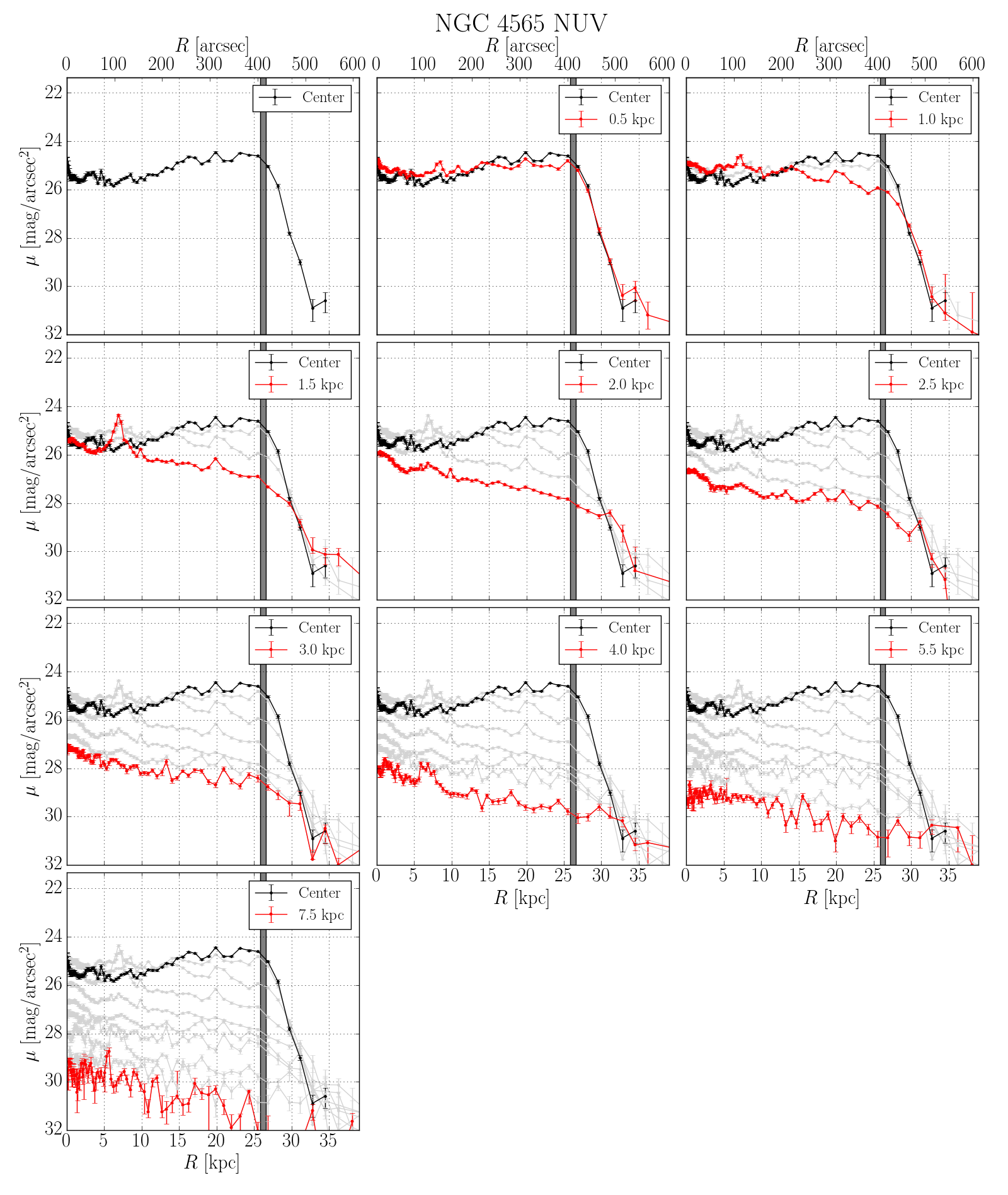}\hspace{6mm}
\caption{Vertical development of the radial surface brightness profiles in the NUV band for NGC~4565. The mean of the truncation position distribution is equal to 26.2~kpc and the standard deviation is 0.4~kpc.}
\label{fig:4565panelNUV}
\end{figure*}

\begin{figure*}
\centering
\includegraphics[width=\textwidth]{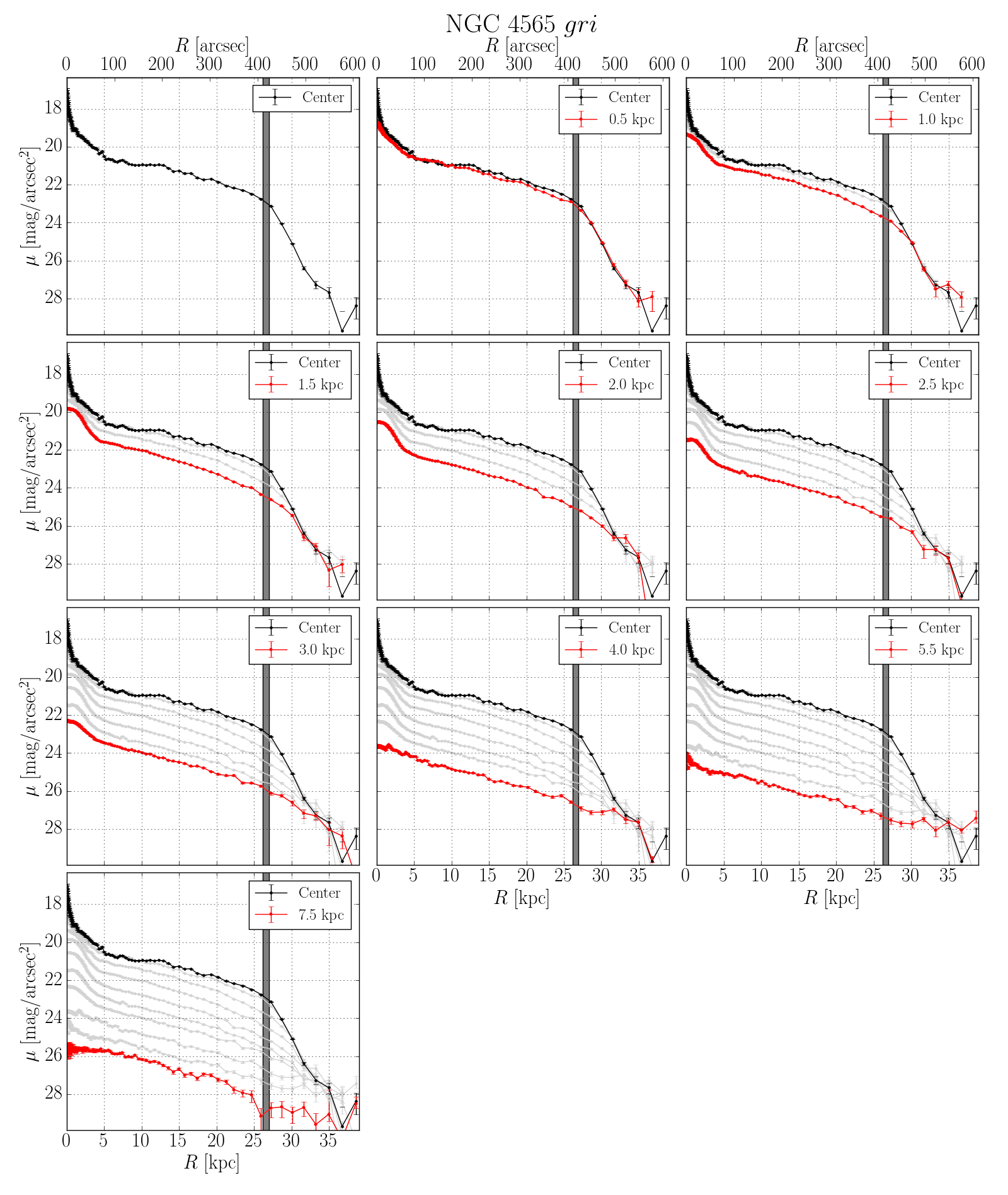}\hspace{6mm}

\caption{As in Fig.~\ref{fig:4565panelNUV}, now for the \textit{gri} combined band. The mean of the truncation position distribution is 26.6~kpc and the standard deviation is 0.4~kpc.}
\label{fig:4565panelgri}
\end{figure*}

\begin{figure*}
\centering
\includegraphics[width=\textwidth]{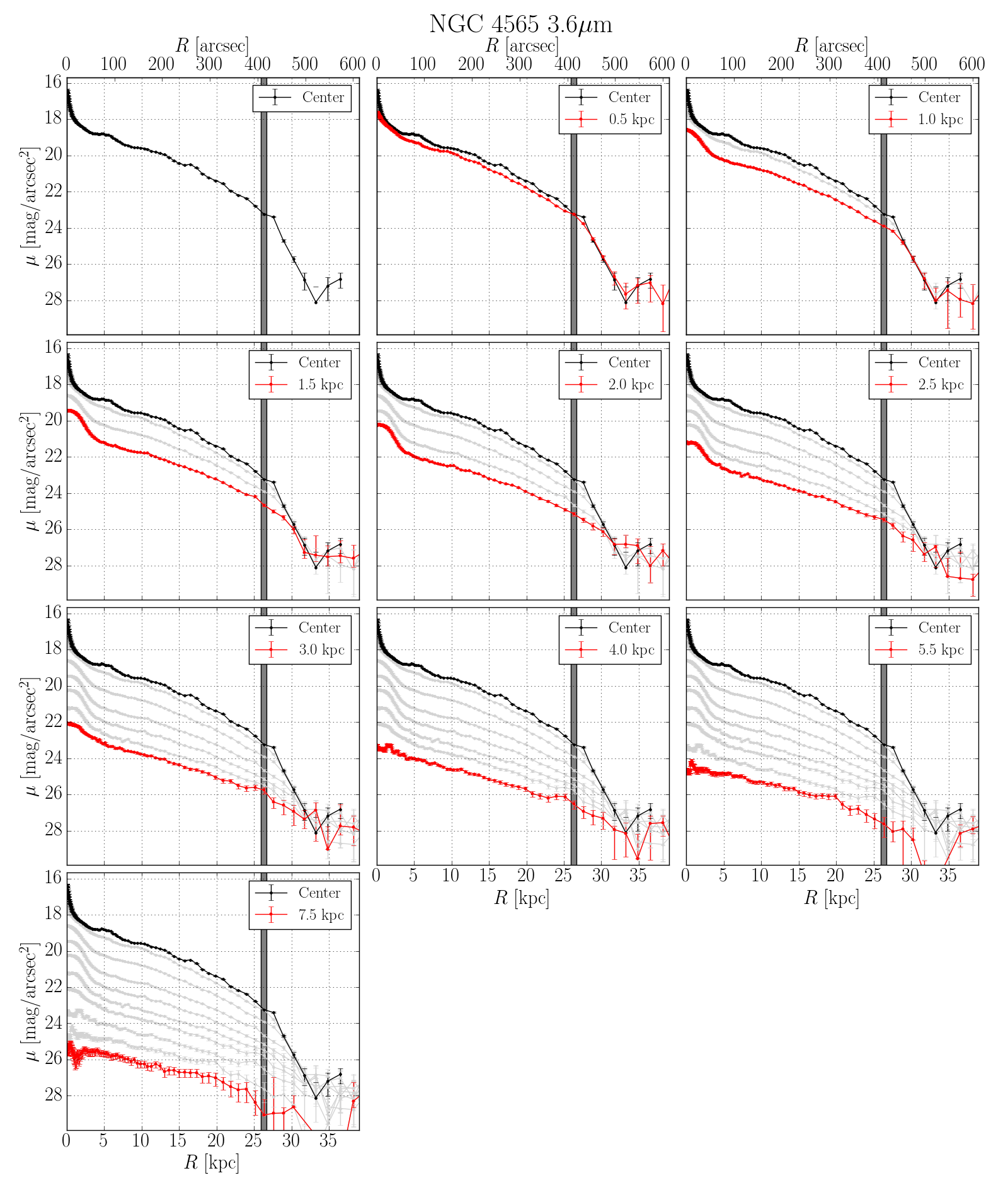}\hspace{6mm}

\caption{As in Fig.~\ref{fig:4565panelNUV}, now in 3.6$\micron$. The mean of the truncation position distribution is 26.3~kpc and the standard deviation is 0.4~kpc.}
\label{fig:4565panel36}
\end{figure*}

\begin{figure*}
\centering

\includegraphics[width=\textwidth]{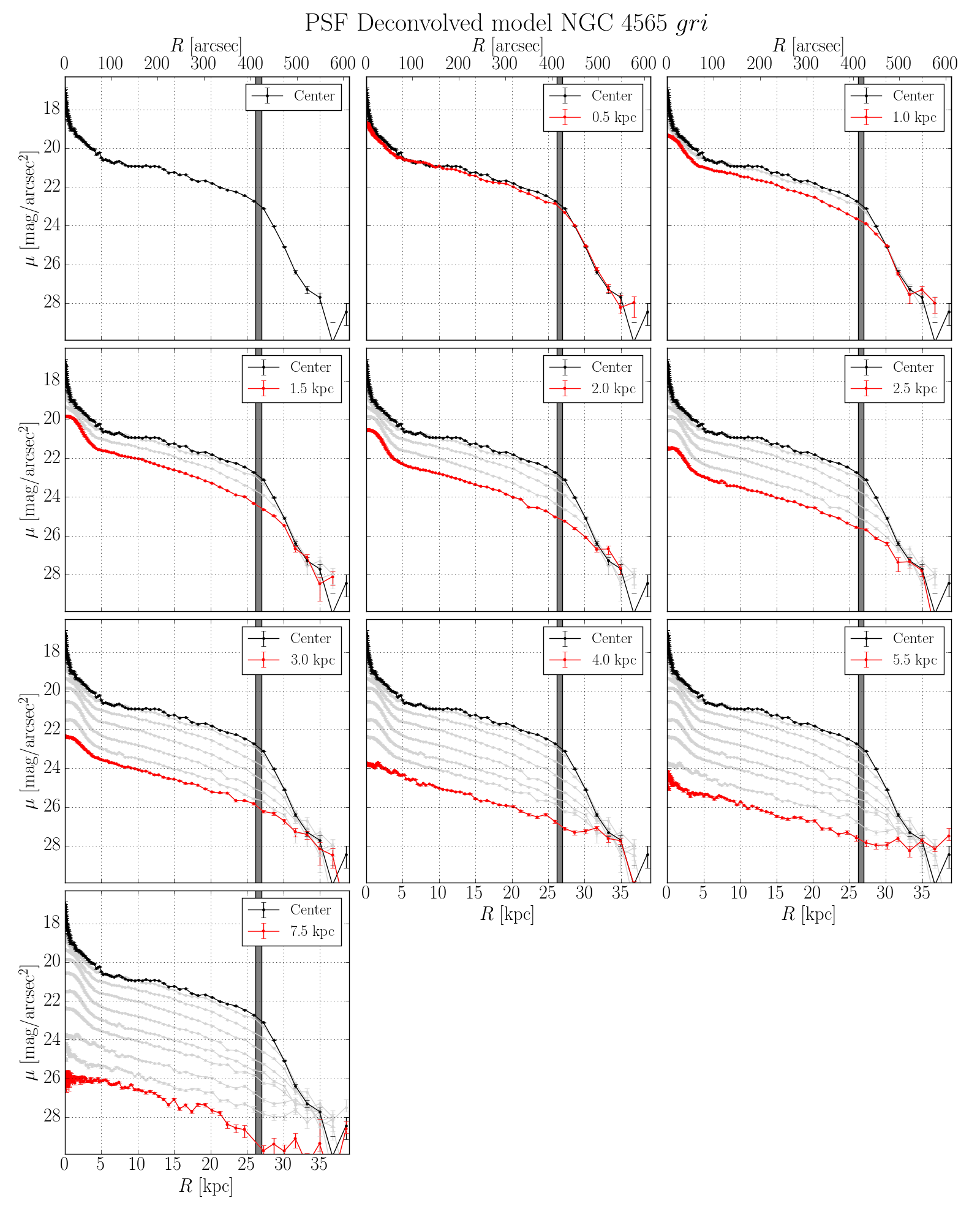}\hspace{6mm}

\caption{As in Fig.~\ref{fig:4565panelNUV}, but for the NGC~4565 PSF-deconvolved model in \textit{gri}. The mean of the truncation position distribution is the same as in \ref{fig:4565panelgri}.}
\label{fig:4565panelModel}
\end{figure*}

\begin{figure*}
\centering
\includegraphics[width=\textwidth]{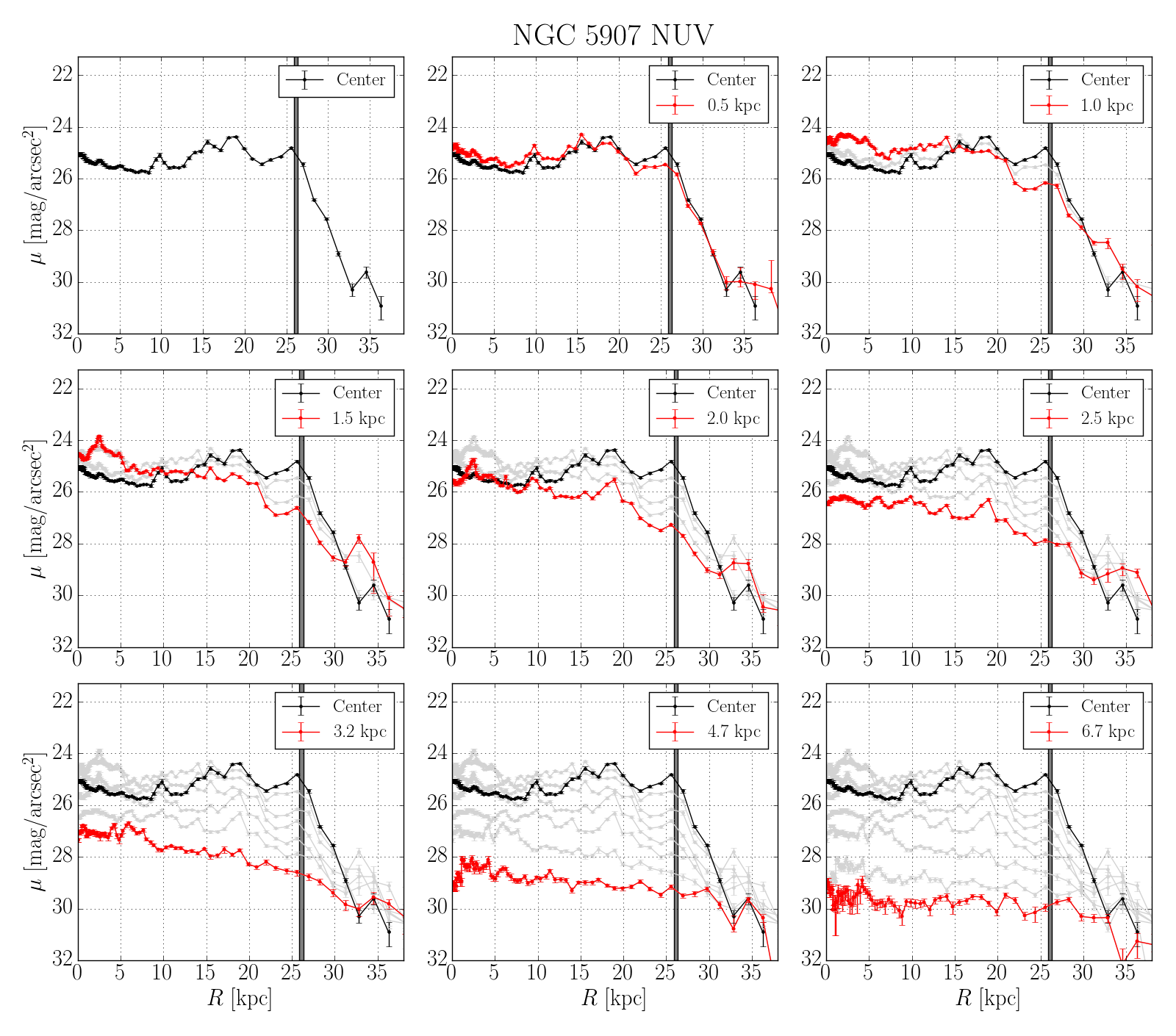}\hspace{6mm}

\caption{As in Fig.~\ref{fig:4565panelNUV}, now for NGC~5907 in NUV band. The mean of the truncation position distribution is 26.1~kpc and the standard deviation is equal to 0.2~kpc.}
\label{fig:5907panelNUV}
\end{figure*}

\begin{figure*}
\centering

\includegraphics[width=\textwidth]{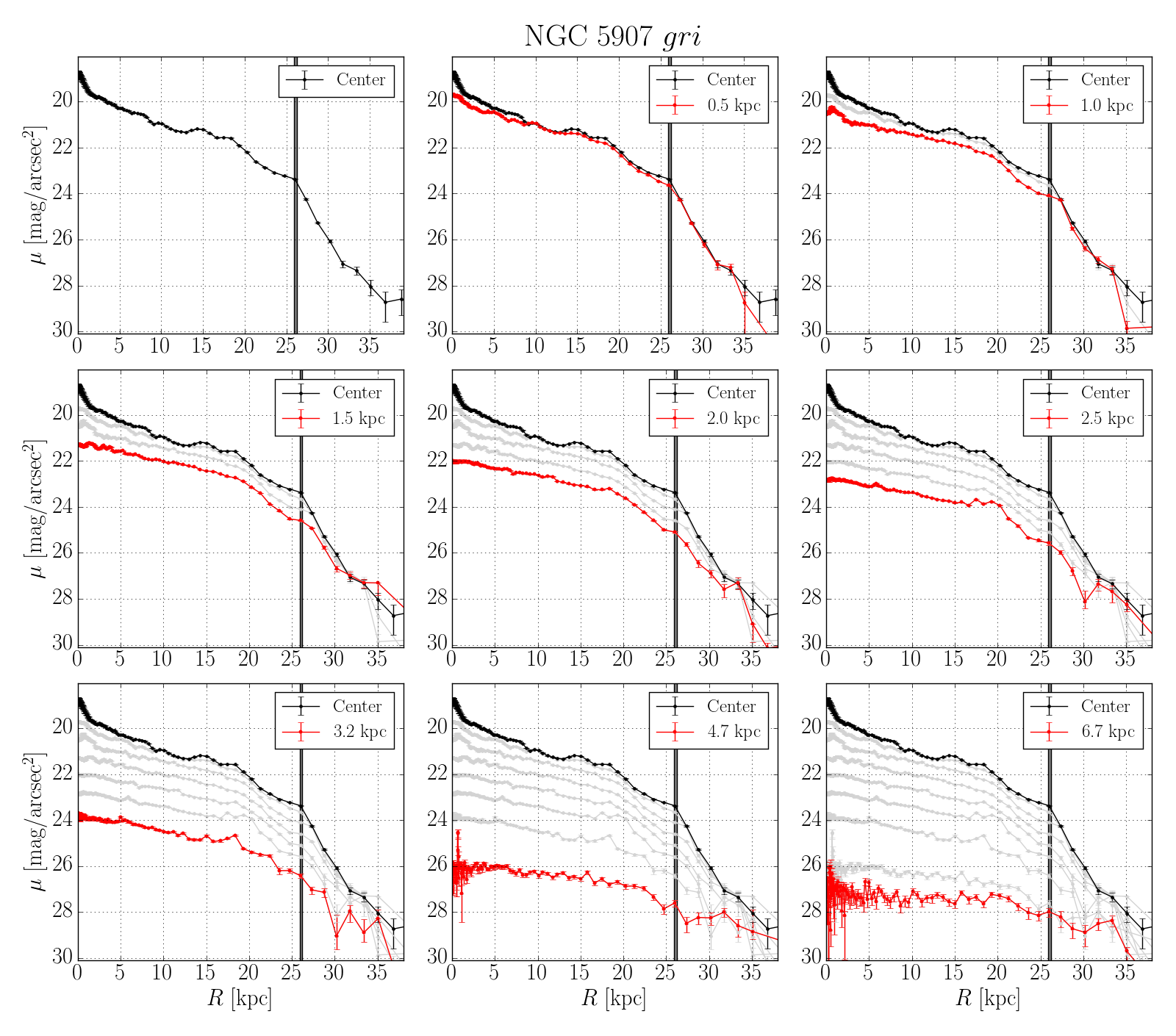}\hspace{6mm}

\caption{As in Fig.~\ref{fig:4565panelNUV}, now for NGC~5907 in the \textit{gri} combined band. The mean of the truncation position distribution is equal to 26.1~kpc and the standard deviation is 0.2~kpc.}
\label{fig:5907panelgri}
\end{figure*}

\begin{figure*}
\centering

\includegraphics[width=\textwidth]{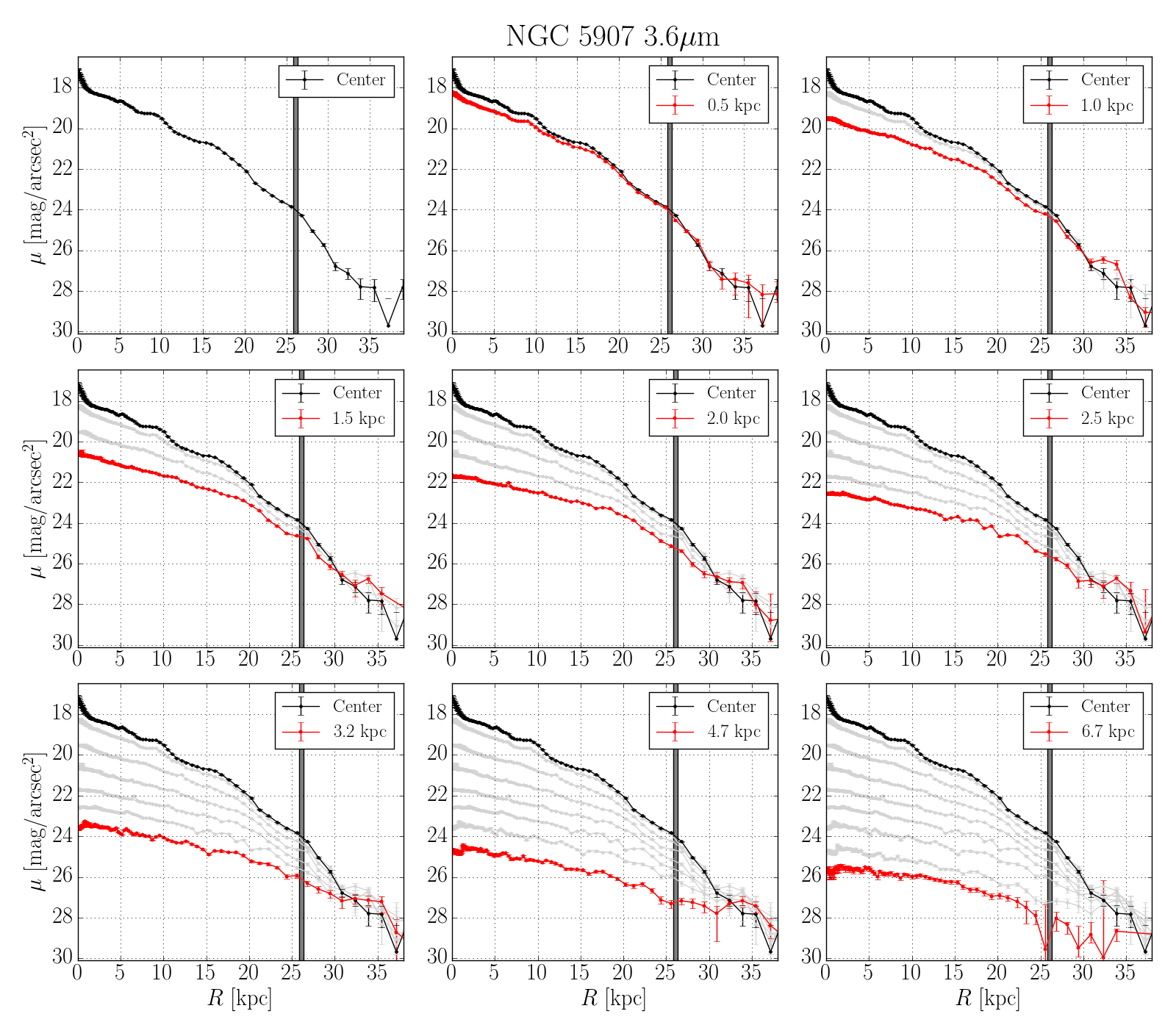}\hspace{6mm}

\caption{As in Fig.~\ref{fig:4565panelNUV}, but for NGC~5907 in 3.6$\micron$. The mean of the truncation position distribution is 26.1~kpc and the standard deviation is 0.3~kpc.}
\label{fig:5907panel36}
\end{figure*}

\begin{figure*}
\centering

\includegraphics[width=\textwidth]{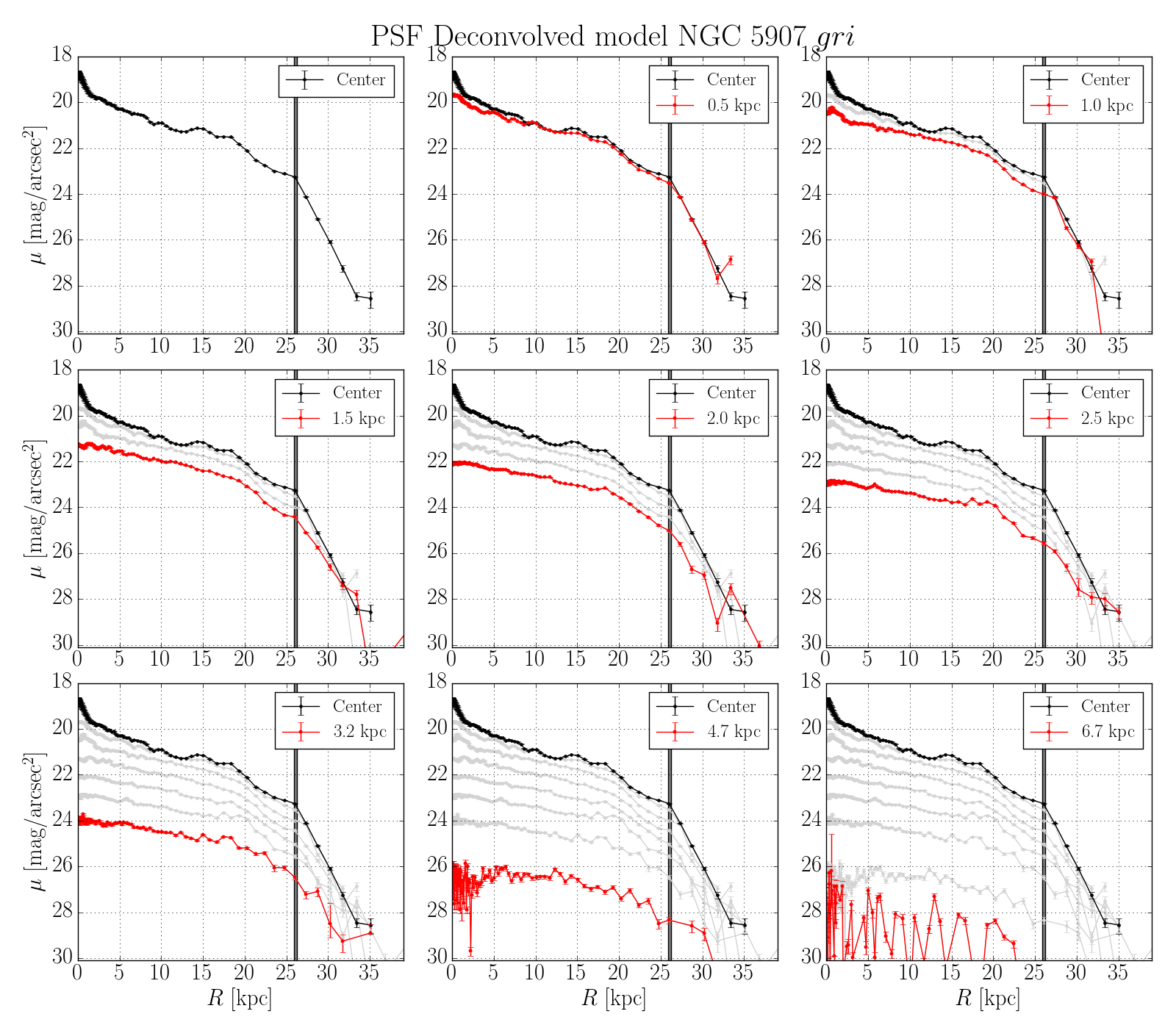}\hspace{6mm}

\caption{As in Fig.~\ref{fig:4565panelNUV}, but for the PSF-deconvolved model of NGC~5907 in \textit{gri}. The mean of the truncation position distribution is the same as in \ref{fig:5907panelgri}.}
\label{fig:5907panelModel}
\end{figure*}

\clearpage

\section{Example of surface brightness profiles above and below the galaxies mid-plane separately} \label{appendixB}

In the next figure, we plot radial surface brightness profiles for each galaxy, in the \textit{gri} band, at three different heights: along the galaxy mid-plane, at 1.5~kpc and at 3~kpc. In each case, we show the surface brightness profiles obtained above the mid-plane of the galaxies (northeastwards, NE), below it (southwesterly, SW) and the mean of both, that is the actual surface brightness profile we show in all the plots of this paper. Here we see that the truncation position is the same in both sides of the galaxies, so using the mean profile in our analysis is a reasonable approach. Note that the main difference is a slight brightness offset between the above and below mid-plane profiles caused by departures from a perfect edge-on inclination of the galaxies. This is more clearly visible in the case of NGC~5907.

\begin{figure*} \label{ApendB}
\includegraphics[width=\textwidth]{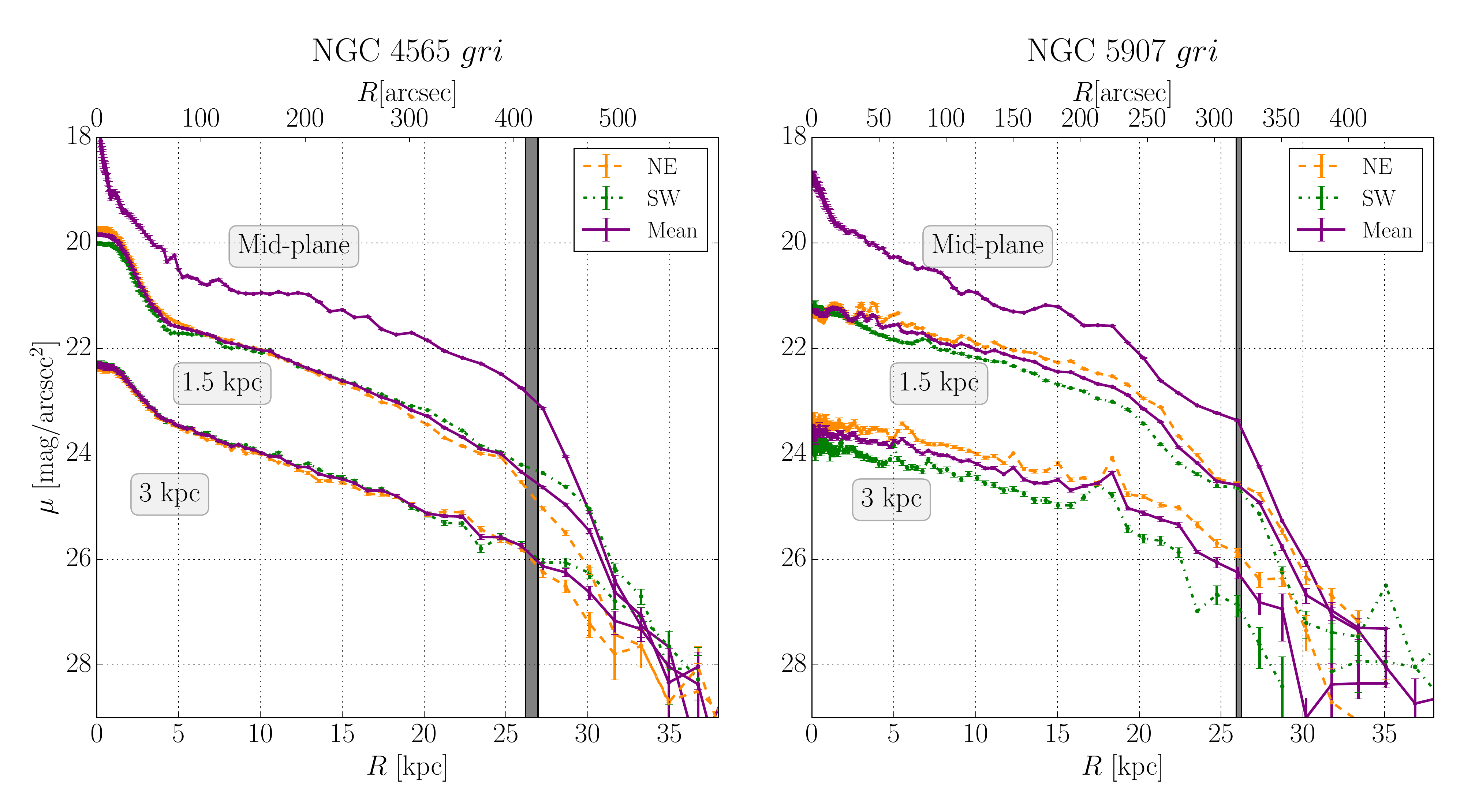}\hspace{6mm}

\caption{Radial surface brightness profiles in the \textit{gri} combined band for both
galaxies along the mid-plane, at 1.5~kpc and 3~kpc above/below the mid-plane. In the two
panels of the figure, the purple solid curves show the mean surface brightness profile at
each height (in the case of the mid-plane profile is the only possibility). The orange
dashed curves represent the profiles above the mid-plane of the galaxies (north-eastwards,
NE) and the green dotted lines are the profiles below the mid-plane (south-westerly, SW).
The height of each set of surface brightness profiles is indicated with the light grey
boxes. The vertical dark grey region always represents the mean position of the truncation
for all the heights in the wavelength range, plus/minus the standard deviation of that
distribution of truncation positions.}

\label{fig:UpDownSum}
\end{figure*}

\clearpage

\clearpage

\section{Example of surface brightness profiles in regions with and without warp} \label{appendixC}

\begin{figure*}
\centering
\includegraphics[width=\textwidth]{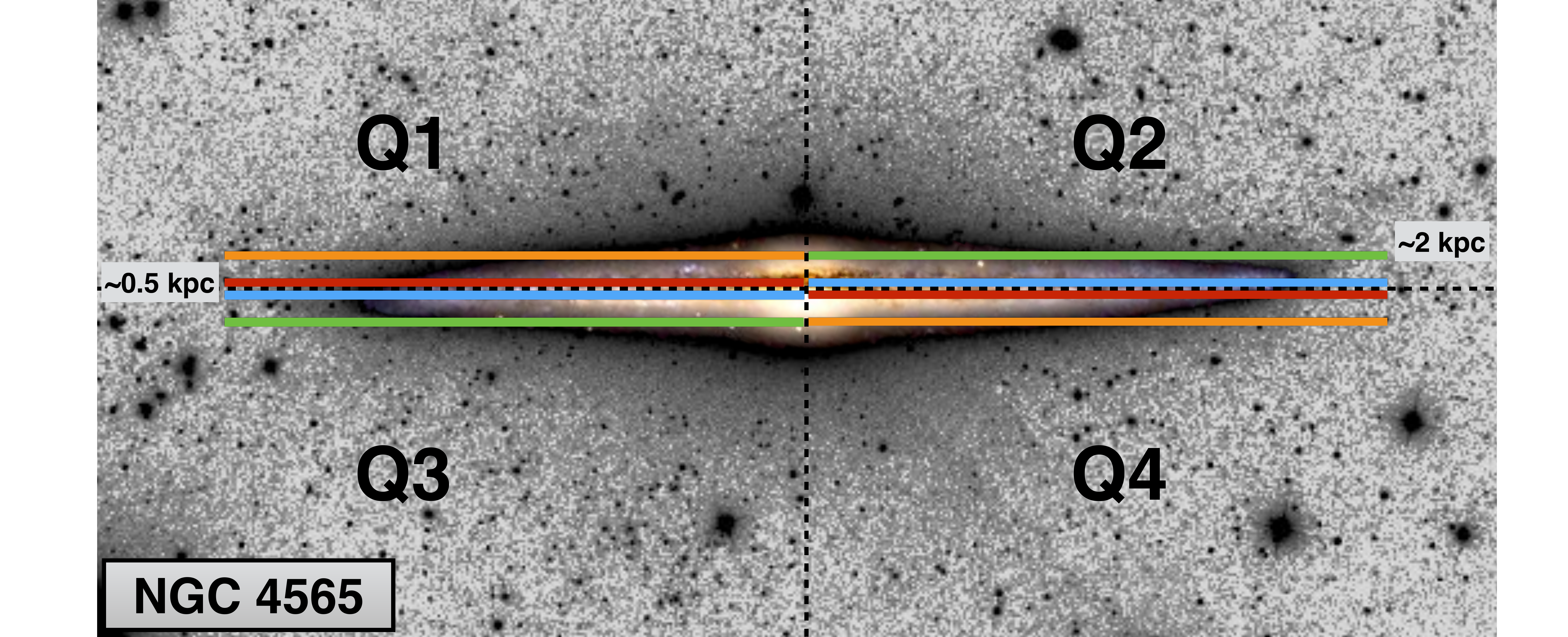}
\caption{Image of NGC~4565 in SDSS \textit{gri} combined light
(roughly equivalent to a deep \textit{r}-band image). For illustrative purposes, a colour image obtained with the same telescope has been inserted atop the saturated disc region of the galaxy. Panels show in greyscale the observed data. The image is split into four quadrants (Q1, Q2, Q3, and Q4). In this galaxy, Q2 and Q3 are the regions with a warp, while Q1 and Q4 are the warp-free areas. The horizontal thick coloured lines indicate the regions where the surface brightness profiles in Fig.~\ref{fig:sbp_warp} were taken. The lines with same colour indicate that the profiles of both regions were averaged in a single surface brightness profile. The profiles closer to the galaxy's mid-plane are at $0.5$~kpc height, while those above are located at $2$~kpc.}
\label{fig:4565_Q}
\end{figure*}

\begin{figure*}
\centering
\includegraphics[width=\textwidth]{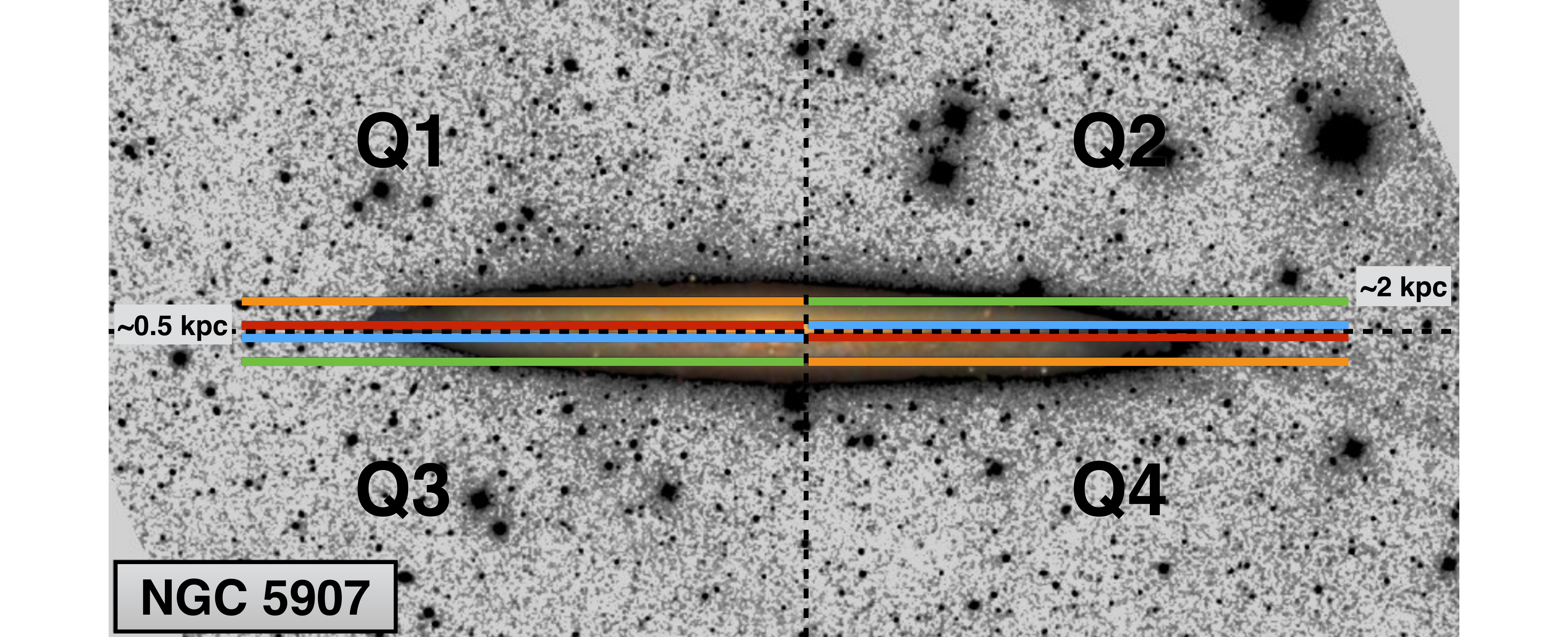}
\caption{As in Fig.~\ref{fig:4565_Q}, but now for NGC~4565. In this galaxy, Q1 and Q4 are the regions with a warp, while Q2 and Q3 are the warp-free areas.}
\label{fig:5907_Q}
\end{figure*}

\begin{figure*}
\centering
\includegraphics[width=\textwidth]{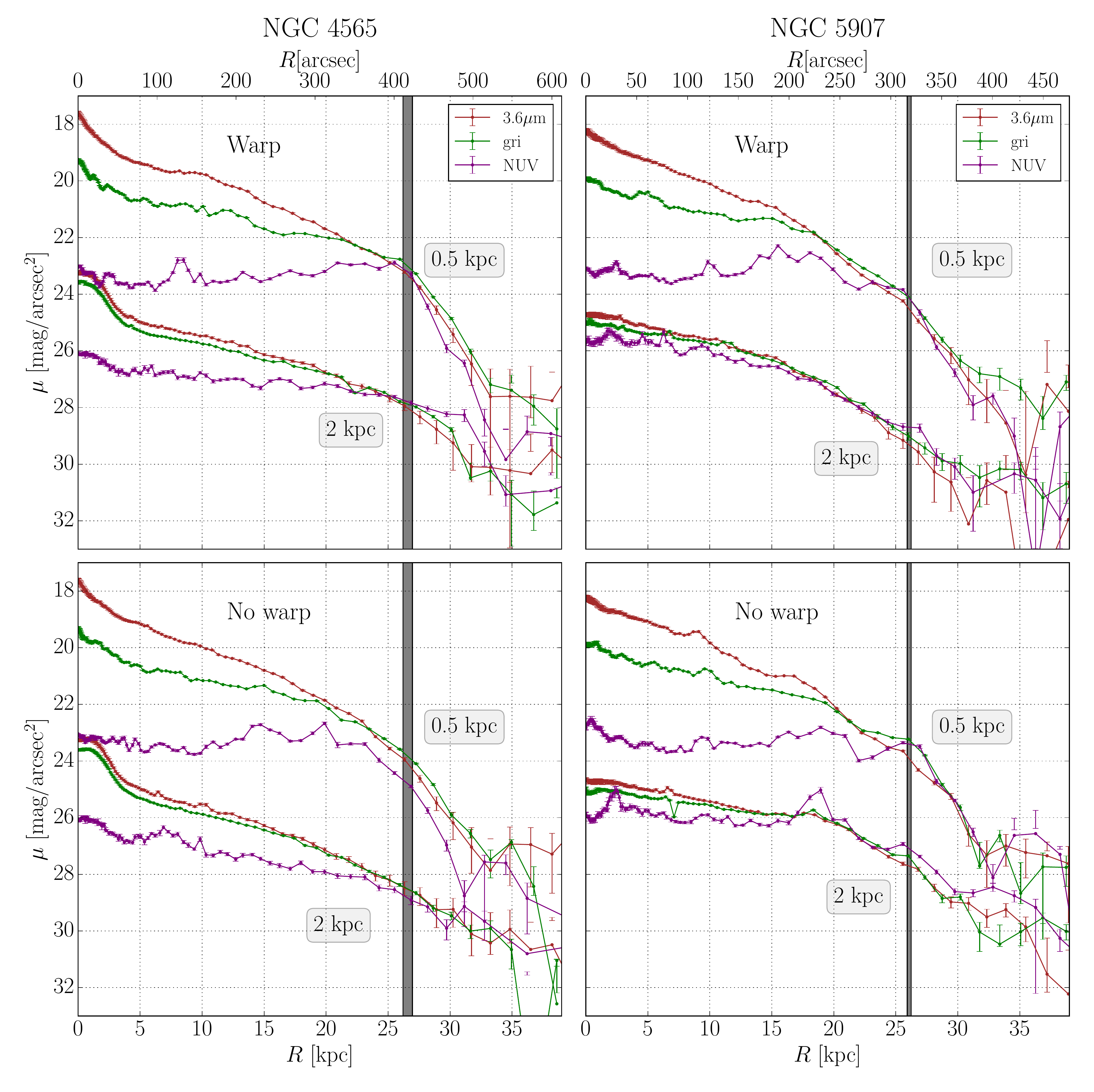}
\caption{Radial surface brightness profiles obtained from the observed data of each galaxy, NGC~4565 (left column) and NGC~5907 (right column), at two different heights above/below the galaxy mid-plane (0.5~kpc and 2~kpc ), in the NUV, \textit{gri}, and 3.6$\micron$ bands, for the warped (top row) and non-warped (bottom row) regions separately. For NGC~4565, the warped region corresponds to quadrants Q2 and Q3 in Fig.~\ref{fig:4565_Q}. Thus, the surface brightness profiles at 0.5~kpc correspond to the blue thick lines, and the ones at 2~kpc, to the green lines shown in Fig.~\ref{fig:4565_Q}. Q1 and Q4 are the non-warped regions in NGC~4565, and the red and orange lines in Fig.~\ref{fig:4565_Q} indicate the locations we used to obtain the 0.5~kpc and  2~kpc surface brightness profiles. We proceed identically for NGC~5907, but considering that Q1 and Q4 are the quadrants with a warp in Fig.~\ref{fig:5907_Q}, so in Q2 and Q3 there is no warp. The surface brightness profiles are obtained using regions with $0.5$~kpc width. The vertical dark grey region represents the mean position of the truncation for all the heights in the SDSS \textit{gri} combined band for each galaxy, plus/minus the standard deviation of that distribution of truncation positions.}
\label{fig:sbp_warp}
\end{figure*}

In Figs.~\ref{fig:4565_Q} and \ref{fig:5907_Q}, we show images in the SDSS \textit{gri} combined band of NGC~4565 and NGC~5907, respectively. We indicate the regions of each galaxy in which the warps are located by splitting the images into four quadrants (Q1, Q2, Q3, and Q4). We then show surface brightness profiles taken in those four quadrants in Fig.~\ref{fig:sbp_warp}.

In Fig.~\ref{fig:sbp_warp}, we plot the radial surface brightness profiles in the warped and non-warped regions for each galaxy, in all wavelength ranges (NUV, \textit{gri}, and 3.6$\micron$), and at two different heights above/below the galaxy mid-plane: $0.5$~kpc and $2$~kpc. The surface brightness profiles are $0.5$~kpc width.   


%
%
%

%
%
%

\clearpage

\bsp	
\label{lastpage}
\end{document}